\documentclass[lettersize,journal]{IEEEtran}
\usepackage{amsmath,amssymb,amsthm,amsfonts,mathrsfs, mathtools, bm, bbm, dsfont, epstopdf,algorithm}
\usepackage{verbatim}
\makeatletter
\let\NAT@parse\undefined
\makeatother
\usepackage[numbers]{natbib}

\usepackage{enumitem}
\usepackage{microtype}
\usepackage{graphicx,caption,subfigure}
\usepackage{csvsimple,booktabs}
\usepackage{textcomp}
\usepackage[dvipsnames]{xcolor}
\usepackage{hyperref}

\newcommand{\beq}{\begin{equation}}
\newcommand{\eeq}{\end{equation}}
\newcommand{\beqa}{\begin{eqnarray}}
\newcommand{\eeqa}{\end{eqnarray}}
\newcommand{\beqan}{\begin{eqnarray*}}
\newcommand{\eeqan}{\end{eqnarray*}}

\newcommand{\rank}{{\rm rank}}

\newcommand{\vnorm}[1]{\left\|#1\right\|}

\newcommand{\dd}{\mathrm{d}}

\newcommand{\Nset}{\mathbb{N}}

\newcommand{\Rset}{\mathbb{R}}

\newcommand{\Xset}{\mathbb{X}}

\newcommand{\Zset}{\mathbb{Z}}

\newcommand{\Bcal}{{\cal B}}
\newcommand{\Ccal}{{\cal C}}

\newcommand{\Fcal}{{\cal F}}

\newcommand{\Hcal}{{\cal H}}

\newcommand{\Kcal}{{\cal K}}
\newcommand{\Lcal}{{\cal L}}
\newcommand{\Mcal}{{\cal M}}

\newcommand{\Ocal}{{\cal O}}

\newcommand{\Tcal}{{\cal T}}
\newcommand{\Ucal}{{\cal U}}
\newcommand{\Vcal}{{\cal V}}
\newcommand{\Wcal}{{\cal W}}

\newcommand{\Ycal}{{\cal Y}}
\newcommand{\Zcal}{{\cal Z}}

\newcommand{\Usf}{\sf{U}}

\newcommand{\Ysf}{\sf{Y}}

\newcounter{l1}
\newcounter{l2}
\newcounter{l3}
\newcommand{\bdotlist}{\begin{list}{$\bullet$}{}}
\newcommand{\bboxlist}{\begin{list}{$\Box$}{}}
\newcommand{\bbboxlist}{\begin{list}{\raisebox{.005in}{{\tiny
$\blacksquare$ \ \ }}}{}}
\newcommand{\bdashlist}{\begin{list}{$-$}{} }
\newcommand{\blist}{\begin{list}{}{} }
\newcommand{\barablist}{\begin{list}{\arabic{l1}}{\usecounter{l1}}}
\newcommand{\balphlist}{\begin{list}{(\alph{l2})}{\usecounter{l2}}}
\newcommand{\bAlphlist}{\begin{list}{\Alph{l2}.}{\usecounter{l2}}}
\newcommand{\bdiamlist}{\begin{list}{$\diamond$}{}}
\newcommand{\bromalist}{\begin{list}{(\roman{l3})}{\usecounter{l3}}}

\newtheorem{theorem}{Theorem}

\newtheorem{lemma}[theorem]{Lemma}

\newtheorem{corollary}[theorem]{Corollary}

\renewcommand{\hat}{\widehat}
\renewcommand{\tilde}{\widetilde}
\newcommand{\obar}{\overline}
\newcommand{\Tr}{{\textrm{Tr}}}
\newcommand{\range}{\text{range}}

\allowdisplaybreaks

\def\trn{{\hbox{\it\tiny T}}} 
\def\data{{\rm{d}}}
\def\future{{\rm{f}}}
\def\past{{\rm{p}}}
\def\s{{\rm{s}}}

\title{\LARGE \bf A Behavioral Framework for Data-Driven Modeling of Nonlinear Systems in Vector-Valued Reproducing Kernel Hilbert Spaces}

\author{Boya Hou$^{1}$ and Maxim Raginsky$^{2}$%
\thanks{This work
was supported in part by the NSF under award CCF-2348624 (''Towards a control framework for neural generative modeling”).}%
\thanks{$^{1}$ Boya Hou is with the Carl R. Woese Institute for Genomic Biology and  the Coordinated Science Laboratory, University of Illinois Urbana-Champaign, Urbana, IL, 61801 USA {\tt\small boyahou2@illinois.edu}}%
\thanks{$^{2}$ Maxim Raginsky is with Department of Electrical and Computer Engineering and the Coordinated Science Laboratory, University of Illinois Urbana-Champaign, Urbana, IL, 61801 USA {\tt\small maxim@illinois.edu}}
}

\begin{document}

\maketitle

\begin{abstract}
We generalize Jan Willems' behavioral approach to a class of discrete-time \emph{nonlinear} systems in a vector-valued reproducing kernel Hilbert space (RKHS). Apart from linear time-invariant systems, this class covers nonlinear systems modeled by Volterra series and their autoregressive variants, as well as systems admitting Hammerstein-type state-space realizations. We apply the proposed framework to the problem of data-driven modeling of such systems, i.e., when simulation or control objectives for an unknown system are carried out without an explicit system identification step. To that end, we link the behavioral approach to two data-driven modeling methods in a vector-valued RKHS: (1) minimum-norm interpolation and (2) subspace identification. 
\end{abstract}

\section{Introduction}
\label{sec.introduction}
The mathematical study of dynamical systems seeks to describe the evolution of a given system over time under its governing laws, initial conditions, and inputs (if any). This evolution can be studied internally through state-space methods or externally via the system's behavior. Pioneered by Jan Willems in a series of papers starting with \cite{willems1986time} (see \cite{willems1991paradigms} for an overview), the behavioral approach identifies dynamical systems with sets of their trajectories. For linear time-invariant systems, the cornerstone of this framework is the so-called \textit{fundamental lemma} \cite{willems2005note}, which asserts that the set of all trajectories of a controllable linear time-invariant system over a finite time horizon can be reconstructed from a single trajectory driven by a persistently exciting input. This result has become central to recent advances of data-driven control; see, e.g., \cite{markovsky2006exact,markovsky2008data,coulson2019data,dorfler2022bridging,van2023informativity} and references therein.

The philosophy underlying these data-driven approaches, as clearly articulated in the paper by Markovsky and Rapisarda \cite{markovsky2008data}, is that simulation or control should be carried out without an explicit system identification step. 
In the context of simulation of linear systems, for example, this principle is supported by the fact that, given an initial condition and a subsequent input of interest, one can combine it with previous measurements of the system behavior to predict the resulting output without identifying the system transfer function, impulse response, or state-space model. Rather, simulation can be viewed as a ``missing data'' problem, to be solved directly using the available trajectories.

While this viewpoint has led to a rich and mature literature for linear systems, extending it to nonlinear systems is substantially more challenging, as the trajectory sets no longer form linear subspaces. Several works have explored nonlinear analogues of the fundamental lemma for structured nonlinear systems. Examples include bilinear systems \cite{markovsky2022data}, second-order Volterra systems \cite{rueda2020data}, flat nonlinear systems \cite{alsalti2021data}, Hammerstein and Wiener systems \cite{berberich2020trajectory}, and a Koopman-type linear embedding posited in \cite{shang2024willems}. A promising direction is to use a reproducing kernel Hilbert space (RKHS) to lift the trajectories of nonlinear systems into a (possibly infinite-dimensional) feature space. In this vein, Huang, Lygeros, and D\"orfler \cite{huang2023robust} investigated kernelized data-enabled predictive control, where the resulting models are regression-based approximations rather than behavioral characterizations. More recently, Molodchyk and Faulwasser \cite{molodchyk2024RKHS} established the link between kernel regression and Willems’ fundamental lemma, and showed that several existing nonlinear extensions of the fundamental lemma, such as those for Hammerstein and flat systems, can be interpreted as instances of kernel regression with specific choices of kernel when the induced RKHS is finite-dimensional.

In this paper, we extend the behavioral framework beyond the linear settings for characterizing the behavior of nonlinear systems in RKHS via minimum-norm interpolation and subspace identification repurposed for nonlinear systems. We reinterpret both minimum-norm interpolation and subspace identification through the behavioral lens, and clarify the role of various structural assumptions on the system, the offline and online data, and the space of functions used to represent the nonlinear aspects of the system. A unifying theme across these two methods is that, like in the modern data-driven schemes and in the spirit of the fundamental lemma, predictors and the desired ``online'' trajectories can be expressed in terms of the kernelized ``offline'' examples. When working in an RKHS, as noted by Molodchyk and Faulwasser \cite{molodchyk2024RKHS}, this structure further mirrors the representer theorem \cite{scholkopf2001generalized}: the solution can be written as a finite linear combination of kernel evaluations along the observed trajectories.

The remainder of this paper is organized as follows.
In Section \ref{sec.preliminary}, we recall the fundamental lemma of Willems for linear time-invariant systems and introduce basic concepts regarding scalar-valued and vector-valued RKHSs. In Section \ref{sec.3}, we introduce a class of nonlinear systems whose nonlinearity is encoded by an element in the vector-valued RKHS of functions on the set of finite-length sequences of inputs. In Section \ref{sec.interpolate}, we establish a \emph{Behavior Representer Theorem} for nonlinear systems using minimum-norm interpolation and contrast it with the fundamental lemma. Compared with the work of Molodchyk and Faulwasser \cite{molodchyk2024RKHS}, we consider minimum-norm interpolation instead of (regularized) least-squares regression in a vector-valued RKHS. We further establish an error estimate that holds with \textit{equality} as in Lemma \ref{lemma.Sigma.t}, rather than the inequaliity in \cite[Lemma 2]{molodchyk2024RKHS}. In Section \ref{sec.subspace}, we present a subspace identification framework for nonlinear systems in an RKHS and provide a fundamental lemma  characterization of finite-length trajectories with kernelized input vectors and outputs.

\section{Preliminaries}
\label{sec.preliminary}

\subsection{Notation and definitions}

We will make use of the following notation and definitions throughout the paper. We will use $\Zset_+$ to denote the set of nonnegative integers. The Moore--Penrose pseudoinverse of a matrix $A$ will be denoted by $A^\dagger$.  Given the matrices $A \in \Rset^{p \times k}$, $B \in \Rset^{q \times k}$, and $C \in \Rset^{r \times k}$, the \textit{oblique projection} of the rowspace of $A$ on the rowspace of $C$ along the rowspace of $B$ is defined as
\begin{align*}
	A \underset{B}{/} C := A \begin{bmatrix} C^\trn & B^\trn \end{bmatrix} \left( \begin{bmatrix} CC^\trn & CB^\trn \\
	BC^\trn & BB^\trn \end{bmatrix}^\dagger \right)_{\text{first $r$ columns}} C
 \end{align*}
(see \cite[Section~1.4.2]{van2012subspace}). Given a finite sequence of vectors $w_{0:T-1} = \{w_t\}^{T-1}_{t=0}$ in $\Rset^q$, the \textit{Hankel matrix of depth $L$} is the $(Lq) \times (T-L+1)$ matrix given by
\begin{align*}
	H_L(w_{0:T-1}) := \begin{bmatrix}
	w_0 & w_1 & \dots & w_{T-L} \\
	w_1 & w_2 & \dots & w_{T-L+1} \\
	\vdots & \vdots & \ddots & \vdots \\
	w_{L-1} & w_L & \dots & w_{T-1} 
\end{bmatrix}.
\end{align*}
We say that $w_{0:T-1}$ is \textit{persistently exciting} (PE) \textit{of order $L$} if the Hankel matrix $H_L(w_{0:T-1})$ has full row rank. Given a signal (discrete-time vector-valued sequence) $\{w_t\}_{t \in \Zset_+}$, $\sigma$ is the backward shift operator defined by $(\sigma w)_t := w_{t+1}$. Additional notation will be introduced as needed.

\subsection{A review of the fundamental lemma}

In the behavioral framework, a linear time-invariant system with $q$ variables is identified with a linear subspace $\Bcal$ of the space of all one-sided sequences of $q$-dimensional real vectors denoted as $(\Rset^q)^{\Zset_+}$ which is shift-invariant, i.e., $\sigma \Bcal \subseteq \Bcal$. One can work with various equivalent \textit{representations} of $\Bcal$, such as autoregressive, input/output, or input/state/output representations \cite{willems1986time}. Notions like controllability or observability can be defined solely in terms of set-theoretic properties of $\Bcal$, without referring to a particular representation \cite{willems1991paradigms}.

One of these structural properties pertains to the partition of the variables of $\Bcal$ into inputs and outputs. Without getting too much into technical details, the idea is that, up to a permutation of coordinates, each trajectory $w \in \Bcal$ can be partitioned into an input trajectory $u : \Zset_+ \to \Rset^{m}$ and an output trajectory $y : \Zset_+ \to \Rset^{p}$ as $w = \begin{bmatrix} u \\ y \end{bmatrix}$, such that the input is free in the sense that
\begin{align*}
	\begin{aligned}
	&\left\{ u : \Zset_+ \to \Rset^{m} \,\middle|\, \begin{bmatrix} u \\ y \end{bmatrix} \in \Bcal \text{ for some } y : \Zset_+ \to \Rset^{p}\right\} \\
	& \qquad \qquad \simeq (\Rset^{m})^{\Zset_+}
	\end{aligned},
\end{align*}
and the output is determined by the input, subject to the system laws and the initial condition. It can be shown that the number of inputs $m$, the number of outputs $p$, and the dimension $n$ of any minimal state space realization of $\Bcal$ are system invariants that depend only on $\Bcal$ and not on the particular representation of $\Bcal$ \cite{willems1986time}.

Let $\Bcal$ be a controllable linear time-invariant system with $m$ inputs, $p$ outputs, and minimum state dimension $n$. The main result of \cite{willems2005note}, now commonly referred to as the \textit{fundamental lemma}, is as follows:

\begin{lemma} For each $t = 1,2,\dots$, let $\Bcal|_t$ denote the restriction of $\Bcal$ to times $s \in \{0,\dots,t-1\}$. Let a trajectory $w^\data_{0:T-1} \in \Bcal|_T$ be given, such that the input part of $w^\data_{0:T-1}$ is persistently exciting of order $L+n$. Then $\Bcal|_L$ is equal to the column space of the Hankel matrix $H_L(w^\data_{0:T-1})$.\label{lm.FL}
\end{lemma}

\noindent The main message of Lemma~\ref{lm.FL} is that the  length-$L$ behavior $\Bcal|_L$ can be reconstructed, exactly and in a representation-independent manner, from a single input/output trajectory of length $T \ge L + (L+n)m - 1$, provided the input is persistently exciting and the system is controllable. This makes the fundamental lemma a key ingredient in data-driven approaches to system simulation and control.
\subsection{Reproducing Kernel Hilbert Spaces}

In this paper, we make extensive use of vector-valued reproducing kernel Hilbert spaces. We start by describing the more familiar definition of a scalar-valued RKHS; we refer interested readers to \cite{berlinet2011reproducing} for details. Let $X$ be a set. A Hilbert space\footnote{Unless indicated otherwise, all Hilbert spaces in this paper are assumed to be defined over the reals.} $\Hcal$ with inner product $\langle \cdot,\cdot \rangle_\Hcal$ is an RKHS on $X$ if its elements are functions from $X$ to $\Rset$ and if for each $x \in X$ there exists a positive constant $C_x$, such that $|f(x)| \le C_x \| f \|_\Hcal$ for all $f \in \Hcal$. In other words, $\Hcal$ is an RKHS on $X$ if the evaluation functional $\delta_x (f) := f(x)$ is bounded (hence continuous) for all $x \in X$. To each RKHS, we can associate a \textit{reproducing kernel}, i.e., a mapping $\kappa : X \times X \rightarrow \Rset$ that satisfies the positivity condition
\begin{align*}
\sum^n_{i,j=1} c_i c_j \kappa(x_i,x_j) \ge 0,
\end{align*}
for all $n \in \Nset$, $c_1,\dots,c_n \in \Rset$, and $x_1,\dots,x_n \in X$, such that $\kappa(\cdot,x)$ is an element of $\Hcal$ for each $x \in X$ and the following property (called the \textit{reproducing kernel property} holds:
\begin{align*}
f(x) = \langle f, \kappa(\cdot,x) \rangle_\Hcal, \qquad \text{for all } f \in \Hcal, x \in X.
\end{align*}
Moreover, the set $\{ \kappa(\cdot,x) : x \in X\}$ is total in $\Hcal$ (i.e., its linear span is dense in $\Hcal$). The map $\phi : X \to \Hcal$ defined by $\phi(x) := \kappa(\cdot,x)$ is called the \textit{canonical feature map}. By the Moore–Aronszajn theorem \cite{aronszajn1950theory}, the reproducing kernel $\kappa$ specifies $\Hcal$ uniquely up to linear isomorphism.

We now describe the vector-valued generalization of the above definition \cite{carmeli_2006}. Let $\Kcal$ be a Hilbert space, and let $\Lcal(\Kcal)$ denote the Hilbert space of bounded linear operators on $\Kcal$ with the Hilbert--Schmidt norm. Then we say that $\Hcal$ is a \textit{$\Kcal$-valued RKHS on $X$} if its elements are functions from $X$ to $\Kcal$ and if the evaluation map $\delta_x (f) := f(x)$ from $\Hcal$ to $\Kcal$ is bounded (hence continuous) for all $x \in X$, i.e., there exists a positive constant $C_x$ such that $\| f(x) \|_\Kcal \le C_x \| f \|_\Hcal$. The vector-valued analogue of the reproducing kernel is an \textit{operator-valued} map $\kappa : X \times X \to \Lcal(\Kcal)$ of the positive type, i.e., such that the inequality
\begin{align*}
\sum^n_{i,j=1} c_ic_j \langle \kappa(x_i,x_j) v,v \rangle_\Kcal \ge 0,
\end{align*}
holds for all $n \in \Nset$, $c_1,\dots,c_n \in \Rset$, $x_1,\dots,x_n \in X$, and $v \in \Kcal$. The reproducing kernel property then takes the form
\begin{align*}
\langle f(x), v \rangle_\Kcal = \langle f,  \kappa(\cdot,x)v \rangle_\Hcal, \text{ for all } f \in \Hcal, x \in X, v \in \Kcal
\end{align*}
where, for each $x \in X$, $\kappa(\cdot,x)$ is an element of the linear space $\Lcal(\Kcal,\Hcal)$ of bounded operators from $\Kcal$ to $\Hcal$. The map $x \mapsto \kappa(\cdot,x)$ from $X$ into $\Lcal(\Kcal,\Hcal)$ is the (operator-valued) canonical feature map.

A basic example of a vector-valued RKHS which will be used frequently in the sequel is $\Hcal = \Lcal(\Vcal,\Wcal)$, where $\Vcal$ and $\Wcal$ are finite-dimensional inner-product spaces and where we equip $\Hcal$ with the Hilbert--Schmidt inner product $\langle A, B \rangle_\Hcal = \Tr (A^*B)$. For the sake of completeness, we present the straightforward proof of the following lemma in Appendix~\ref{ap.LVW}.

\begin{lemma}
$\Hcal$ is a $\Wcal$-valued reproducing kernel Hilbert space on $\Vcal$ with the operator-valued reproducing kernel
$\kappa\left( v,v'\right)
= \langle v,v' \rangle_{\Vcal} I_\Wcal$,
\label{lemma.LVW}
where $I_\Wcal$ is the identity operator on $\Wcal$.
\end{lemma}


\section{Nonlinear Systems in a Vector-Valued RKHS}
\label{sec.3}

We now introduce a class of discrete-time nonlinear systems, where the nonlinearity is represented by an element of a given vector-valued RKHS of functions on the set of finite-length sequences of inputs. We will use the following notation throughout:
\begin{itemize}
\item $\Ucal := \Rset^m$ is the space of inputs;
\item $\Ycal := \Rset^p$ is the space of outputs;
\item ${\Usf} := \Ucal^{L+1}$ is the space of length-$(L+1)$ input sequences, where $L \in \Zset_+$ is a fixed lag;
\item ${\Ysf} := \Ycal^L$ is the space of length-$L$ output sequences;
\item $\Zcal := \Usf \times \Ysf$.
\end{itemize}
 We introduce the system model in Section~\ref{sec.NL.AR} and the associated behavioral constructs in Section~\ref{sec.behaviors}, and close by discussing several examples in Section~\ref{sec.examples}.

\subsection{System model}
\label{sec.NL.AR}

Let $\Hcal_{\Usf}$ be a vector-valued RKHS of functions from $\Usf$ into $\Ycal$, with the operator-valued kernel $\kappa_{\Usf}$ of positive type.\footnote{Here and elsewhere, all finite-dimensional vector spaces are automatically treated as Hilbert spaces with the usual Euclidean inner product.} We consider systems parametrized by $(L+1)$-tuples $(A_1,\dots,A_L,g)$, where $A_1, \dots, A_L \in \Lcal(\Ycal)$ and $g \in \Hcal_{\Usf}$.  The input/output relation corresponding to $(A_1,\dots,A_L,g)$ is given by
\begin{align}
	y_{t+L} + \sum^{L}_{k=1} A_k y_{t+L-k} = g \left( u_{t:t+L} \right), \qquad  t \in \Zset_+
	\label{eq.NL.AR}
\end{align}
where $u_{t:t+L}$ is the restriction of the input sequence $u \in \Ucal^{\Zset_+}$ to the set $\{t,t+1,\dots,t+L\}$. By the reproducing kernel property, we can also write
\begin{align*}
g(u_{t:t+L}) = \kappa_{\Usf}(\cdot, u_{t:t+L})^*g,
\end{align*}
where  $\kappa_{\Usf}\left(\cdot, u_{t:t+L} \right)^* \in \Lcal(\Hcal_{\Usf},\Ycal)$ is the adjoint of the canonical feature map $\kappa_{\Usf}\left(\cdot, u_{t:t+L} \right) \in \Lcal(\Ycal,\Hcal_{\Usf})$.

It is also convenient to cast \eqref{eq.NL.AR} in an alternative nonlinear regression form. For each $t$, let $y_{t^+}$ denote the output $y_{t+L}$ at time $t+L$ and define the \textit{regression vectors}
\begin{align}\label{eq:regression_vectors}
z_t:=[u_{t}^\trn,\cdots, u_{t+L}^\trn, y_{t}^\trn, \cdots y_{t+L-1}^\trn]^\trn \in \Zcal,
\end{align}
which is in analogy to the corresponding construct for linear systems \cite{moore1983PE,Green1986PE}.
As we now show, we can introduce a $\Ycal$-valued RKHS $\Hcal_{\Zcal}$ of functions from $\Zcal$ into $\Ycal$, such that the autoregressive nonlinear model of \eqref{eq.NL.AR} can be represented as
\begin{align}\label{eq.NL.AR.f}
y_{t^+} = f(z_t),
\end{align}
for some $f \in \Hcal_{\Zcal}$ that depends on $(A_1,\dots,A_L,g)$.

To that end, let us first express \eqref{eq.NL.AR} as 
\begin{align}\label{eq.NL.AR.reg}
\begin{aligned}
y_{t+L} &= A^- y_{t:t+L-1} + g\left( u_{t:t+L} \right)\\
&=: f \left(u_{t:t+L} , y_{t:t+L-1}\right),
\end{aligned}
\end{align}
where $A^- \in \Lcal({\Ysf},\Ycal)$ is the linear operator defined by
\begin{align*}
A^- \obar{y} := - \begin{bmatrix} A_1 & A_2 & \dots & A_{L} \end{bmatrix}\obar{y}, \qquad \obar{y} \in {\Ysf}.
\end{align*}
By Lemma~\ref{lemma.LVW}, $\Hcal_{\Ysf} = \Lcal({\Ysf},\Ycal)$ is a $\Ycal$-valued RKHS on ${\Ysf}$ with the reproducing kernel $\kappa_{\Ysf}\left(\obar{y}, \obar{y}'\right)
= (\obar{y}^\trn \obar{y}') I_{\Ycal}$. Consider the direct-sum Hilbert space $\Hcal_{\Zcal} = \Hcal_{\Usf} \oplus \Hcal_{\Ysf}$, whose elements $f: {\Zcal} \to \Ycal$ are pairs $(g,h)$ where $g \in \Hcal_{\Usf}$, $h \in \Hcal_{\Ysf}$, and for each $z = (\obar{u},\obar{y}) \in \Zcal$, $f(z) = g(\obar{u})+h(\obar{y})$. The inner product in $\Hcal_{\Zcal}$ is 
\begin{align*}
\left \langle f_1, f_2 \right \rangle_{\Hcal_{\Zcal}}
= \left \langle g_1,g_2 \right \rangle_{\Hcal_{\Usf}}
+ \left \langle h_1, h_2 \right \rangle_{\Hcal_{\Ysf}},
\end{align*}
for $f_1 = (g_1,h_1)$, $f_2 = (g_2,h_2)$. Since $\Hcal_{\Usf}$, $\Hcal_{\Ysf}$ are $\Ycal$-valued RKHSs with operator-valued kernels $\kappa_{\Usf}$ and $\kappa_{\Ysf}$, $\Hcal_{\Zcal}$ is also an RKHS with operator-valued kernel defined by
\begin{align*}
\begin{split}
\kappa_{\Zcal}(z_1,z_2) &= \kappa_{\Usf}(\obar{u}_1,\obar{u}_2) +  \kappa_{\Ysf}\left(\obar{y}_1, \obar{y}_2 \right), \\
& \qquad z_1 = \left( \obar{u}_1, \obar{y}_1\right), z_2 = \left( \obar{u}_2, \obar{y}_2\right) \in \Zcal.
\end{split}
\end{align*}
This can be proved as follows: for any $f \in \Hcal_{\Zcal}$ and $v \in \Ycal$, we have
\begin{align*}
\begin{aligned}
\left \langle f(z), v \right \rangle_{\Ycal}
=& \left \langle g(z), v \right \rangle_{\Ycal}
+ \left \langle h(z), v \right \rangle_{\Ycal} \\
=& \left \langle g, \kappa_{{\Usf}}(\cdot,\obar{u}) v \right \rangle_{\Hcal_{\Usf}}
+ \left \langle h, \kappa_{{\Ysf}}(\cdot,\obar{y}) v \right \rangle_{\Hcal_{\Ysf}} \\
=& \left \langle f, \left(\kappa_{{\Usf}}(\cdot,\obar{u}) + \kappa_{{\Ysf}}(\cdot,\obar{y}) \right) v \right \rangle_{\Hcal_{\Zcal}},
\end{aligned}
\end{align*}
where the second line follows from the reproducing kernel property. In particular, \eqref{eq.NL.AR.f} holds with the function $f$ defined in \eqref{eq.NL.AR.reg}. 

\subsection{Behaviors}
\label{sec.behaviors}

In this section, we introduce several behavioral constructs related to the system model of Section~\ref{sec.NL.AR}.

Given $(A_1,\dots,A_L,g)$, define the operator $P(\sigma) := A_0\sigma^{L}+ \sum^{L}_{k=1} A_{k} \sigma^{L-k}$ (with $A_0 = I_\Ycal$), where $\sigma$ is the shift operator acting on output sequences. Then \eqref{eq.NL.AR} is equivalent to 
\begin{align}\label{eq:NL.AR.Poly}
\left(P(\sigma) y \right)(t) 
=\kappa_{\Usf}\left(\cdot, u_{t:t+L} \right)^* g, \qquad t \in \Zset_+,
\end{align}
and we can define the behavior of $(A_1,\dots,A_L,g)$ as the following subset of the space of all input/output sequences:
\begin{align}
\begin{aligned}
&\Bcal(A_1,\dots,A_L,g) \\
&\equiv \Bcal(P,g) \\
&:=\left\{
\begin{bmatrix} u \\ y \end{bmatrix} \in (\Ucal \times \Ycal)^{\Zset_+}
\,\middle|\, \eqref{eq:NL.AR.Poly} \text{ holds for all } t \in \Zset_+
 \right\}.
\end{aligned}
\label{def.B.NL-AR}
\end{align}
For $t \in \Zset_+$, the restriction of the behavior $\Bcal(P,g)$ to the finite time interval  $\{t,\dots,t+{L}\}$ is defined by
\begin{align*}
\begin{split}
\Bcal(P,g)|_{t:t+{L}} = \left\{
\begin{bmatrix} u_{t:t+L} \\ y_{t:t+L} \end{bmatrix} \,\middle|\, 
\exists \begin{bmatrix} \bar{u} \\ \bar{y} \end{bmatrix} \in \Bcal(P,g) \right. \\
\left.\text{ such that } \begin{bmatrix} \bar{u}_{t:t+{L}} \\ \bar{y}_{t:t+{L}} \end{bmatrix} 
= \begin{bmatrix} u_{t:t+L} \\ y_{t:t+L} \end{bmatrix}
\right\}.
\end{split}
\end{align*}
When $t = 0$, we will use $\Bcal(P,g)|_{L}$ as shorthand for $\Bcal(P,g)|_{t:t+{L}}$. 

Next, we turn to the nonlinear regression form in \eqref{eq.NL.AR.f}. Recall that the identity
\begin{align*}
f(z) = \kappa_{\Zcal}(\cdot,z)^*f
\end{align*}
holds by the reproducing kernel property. Define the operator $R : (\Ucal \times \Ycal)^{\Zset_+} \to \Lcal(\Hcal_\Zcal,\Ycal) \times \Ycal $ by
\begin{align*}
R \begin{bmatrix} u \\ y \end{bmatrix} := (\kappa_{\Zcal}(\cdot,z_0)^*, y_{0^+}),
\end{align*}
where
\begin{align}\label{eq.z0_y0plus}
z_0 =[u_{0}^\trn,\cdots, u_{L}^\trn, y_{0}^\trn, \cdots y_{L-1}^\trn]^\trn, \quad y_{0^+} = y_L,
\end{align}
is the regression vector at time $t = 0$ [cf.~Eq.~\eqref{eq:regression_vectors}] and $y_{0^+}$ is the output at time $t = L$. We say that
 $(H, y) \in \Lcal\left(\Hcal_\Zcal,\Ycal \right) \times \Ycal $ is an input-output (i/o) pair of the system  \eqref{eq.NL.AR.f} if $Hf = y$. This 
 leads naturally to the following behavioral description:
\begin{align*}
\begin{aligned}
\Bcal(f) 
&:=\Bigg\{
\begin{bmatrix} u \\ y \end{bmatrix} \in (\Ucal \times \Ycal)^{\Zset_+}
\Bigg| R \sigma^t \begin{bmatrix} u \\ y \end{bmatrix} \\
& \quad \text{ is an i/o pair of \eqref{eq.NL.AR.f} for each $t \in \Zset_+$} \Bigg\}.
\end{aligned}
\end{align*}
When $(A_1,\dots,A_L,g)$ and $f$ are related via \eqref{eq.NL.AR.f}, it is easily verified that
\begin{align}
\label{eq.behavior.equivalence}
\Bcal(A_1,\dots,A_L,g)|_L = \left\{
\begin{bmatrix} u_{0:L} \\ y_{0:L} \end{bmatrix} \,\middle|\, \kappa_\Zcal(\cdot,z_0)^* f = y_{0^+}
\right\},
\end{align}
where $z_0$ and $y_{0^+}$ are defined in \eqref{eq.z0_y0plus}.

\subsection{Examples}
\label{sec.examples}

\subsubsection{LTI Systems} Let $\Hcal_{\Usf}$ consist of all linear operators from $\Usf$ into $\Ycal$, i.e., $\Hcal_{\Usf} = \Lcal(\Usf,\Ycal)$. By Lemma~\ref{lemma.LVW}, this is a $\Ycal$-valued RKHS on $\Usf$. Any $g \in \Hcal_{\Usf}$ can be represented as $g(u_{0:L}) = B_L u_0 + \cdots + B_0 u_{L}$ for some linear operators $B_0,\dots,B_L \in \Lcal(\Ucal,\Ycal)$.
Then Eq.~\eqref{eq.NL.AR}  describes a linear autoregressive model of order ${L}$:
\begin{align*}
y_{t+L} + \sum^L_{k=1} A_k y_{t+L-k} = \sum^L_{l = 0} B_{l} u_{t+L-l}.
\end{align*}
The operator-valued kernel map is given by
\begin{align*}
\kappa_{\Usf}(u_{0:L},v_{0:L}) = \langle u_{0:L}, v_{0:L} \rangle_{\Usf} I_\Ycal,
\end{align*}
where 
\begin{align*}
 \langle u_{0:L}, v_{0:L} \rangle_{\Usf} = \sum^L_{l = 0} u_l^\trn v_l,
\end{align*}
is the $l^2$ inner product on ${\Usf}$ viewed as a direct sum of $L+1$ Hilbert spaces $\Ucal = \Rset^m$ with the Euclidean inner product $\langle u,v \rangle = u^\trn v$.

\subsubsection{Volterra series}
When $A_k=0$ for all $k = 1,\cdots, {L}$, the nonlinear autoregressive model in \eqref{eq.NL.AR} reduces to
\begin{align}
y_{t+L} = g(u_{t:t+L}), \quad \forall t \in \Nset.
\label{eq.NL.AR.DeF}
\end{align}
This system has finite memory of length $L+1$, since the output at each time $t \ge L$ is determined by the inputs in the finite time window $\{t,t-1,\dots,t-L\}$ of length $L+1$. Systems of this type are often represented using Volterra series. De Figueiredo and Dwyer \cite{de1980best} used the formalism of weighted Fock spaces to connect Volterra series representation to the RKHS framework.

Let us assume for simplicity that $p = 1$ (i.e., the outputs $y_t$ are scalar). Let $\rho = (\rho_k)_{k \ge 0}$ be a sequence of positive reals, such that the infinite series
\begin{align*}
q(\lambda) = \sum^\infty_{k=0} \frac{1}{k!\rho_k} \lambda^k,
\end{align*}
converges for all $\lambda \in \Rset$, and let $\Hcal_{\Usf}$ denote the RKHS of functions from $\Usf$ into $\Rset$ with the reproducing kernel
\begin{align*}
\kappa_{\Usf}(u_{0:L},v_{0:L}) := q(\langle u_{0:L}, v_{0:L} \rangle_{{\Usf}}).
\end{align*}
Let $h_0$ and $(h_k(i_1,\dots,i_k): k \in \Nset, i_1 ,\dots, i_k \in \{0,\dots,L\})$ be a collection of real coefficients satisfying the condition
\begin{align*}
\sum^\infty_{k=0} \frac{\rho_k}{k!} \|h_k\|^2_k < \infty,
\end{align*}
where $\| h_0 \|_0 := |h_0|$ and
\begin{align*}
\| h_k \|_k := \left( \sum^L_{i_1 = 0} \dots \sum^L_{i_k = 1} |h_k(i_1,\dots,i_k)|^2 \right)^{1/2}
\end{align*}
for $k = 1,2,\dots$.  Then any function $g : \Usf \to \Rset$ of the form
\begin{align*}
g(u_{0:L}) = h_0 + \sum^\infty_{k=1} \sum_{i_1,\dots,i_k \in \{0,\dots,L\}} h_k(i_1,\dots,i_k) \prod^k_{j=1} u_{i_j}
\end{align*}
is an element of $\Hcal_{\Usf}$ \cite{de1980best}. 

Removing the restriction $A_1 = \dots = A_L = 0$ yields a broader class of nonlinear systems that subsumes the models studied in \cite{de1980best}.

\subsubsection{State-space models}
\label{sec.NL.ss}
Consider a Hammerstein state-space model of the form
\begin{align}
\begin{aligned}
x_{t+1} &= A x_{t} + B \psi_1\left(u_{t}\right),\\
y_{t} &= C x_{t} + D \psi_2 \left(u_{t}\right),
\end{aligned}
\label{eq.NL.state-space}
\end{align}
where $\psi_1,\psi_2 :\Ucal \to \Rset^q$ are two (nonlinear) functions. We assume that the state $x_t$ takes values in $\Rset^n$, so  $A \in \Rset^{n \times n}$, $B \in \Rset^{n \times q}$, $C \in \Rset^{p \times n}$, and $D \in \Rset^{p \times q}$. We can obtain an input-output representation \eqref{eq.NL.AR} from the state-space representation \eqref{eq.NL.state-space} as follows.

Introduce the Markov parameters $M_j := C A^j B \in \Rset^{p \times q}$ for $j \geq 0$.
Define the $L$-step controllablity matrix $\Ccal_L$, the $L$-step observability matrix $\Ocal_L$, the reversed $L$-step controllability matrix $\Delta_L$, and the modified block Toeplitz operator $\tilde{\Tcal}_L$ as 
\begin{align}
\begin{aligned}
\Ccal_L:=&\begin{bmatrix} B & AB & A^2B & \cdots & A^{{L-1}}B \end{bmatrix}, \\
\Ocal_L:=&\begin{bmatrix} C\\ C A \\ \vdots\\ CA^{{L-1}} 
\end{bmatrix}, \\
\Delta_L:=&\begin{bmatrix} A^{{L-1}}B & \cdots & AB & B  \end{bmatrix},\\
\tilde{\Tcal}_L=& \begin{bmatrix} 
{0} & {0} & \cdots & {0} & {0}\\
M_0 & {0} & \cdots & {0} & {0}\\ 
M_1 & M_0 & \cdots & {0} & {0} \\ 
\vdots & \vdots & \cdots  & \vdots & \vdots \\ 
M_{L-2} & M_{L-2} &  \cdots & M_0 & {0} \end{bmatrix}.
\end{aligned}
\label{def.observablity}
\end{align}
Let us assume that $L \ge n$ is such that $\rank(\Ocal_L) = n$.   Then from \eqref{eq.NL.state-space}, for $k = 0,\cdots,{L}$, we have
\begin{align}\label{eq:Y_k_equations}
\begin{aligned}
y_{t-L+k} =& C A^k x_{t-L} + \sum_{j=0}^{k-1} M_j \psi_1(u_{t-L+k-1-j}) \\
&+ D \psi_2 (u_{t-L+k}).
\end{aligned}
\end{align}
Define the vectors 
\begin{align*}
Y_{t} &:= \begin{bmatrix} y_{t-L} \\ \vdots \\ y_{t-1}\end{bmatrix} \in \Rset^{pL},\\ 
E_{t} &:= \begin{bmatrix} \psi_1(u_{t-L}) \\ \vdots \\ \psi_1(u_{t-1})\end{bmatrix} \in \Rset^{qL}, \\
F_{t} &:=\begin{bmatrix} \psi_2(u_{t-L}) \\ \vdots \\ \psi_2(u_{t-1})\end{bmatrix} \in \Rset^{qL}.
\end{align*}
Then, stacking the equations in \eqref{eq:Y_k_equations} for $k = 0$ to $k = L - 1$, we have
\begin{align*}
Y_{t}
= \Ocal_L x_{t-L} + \tilde{\Tcal}_L E_{t} + \left( I_L \otimes D \right) F_{t}.
\end{align*}
Since $\rank(\Ocal_L) = n$, we can solve for $x_{t-L}$:
\begin{align}
x_{t-L} = \Ocal_L^{\dagger} \left( Y_{t}
- \tilde{\Tcal}_L E_{t} - \left( I_L \otimes D \right) F_{t}\right).
\label{eq.x_{t-L}}
\end{align}
On the other hand, when $k={L}$, we have
\begin{align*}
\begin{aligned}
y_{t} &= C A^L x_{t-L} + \sum_{j=0}^{{L-1}} M_j \psi_1(u_{t-1-j})+ D \psi_2 (u_{t}).
\end{aligned}
\end{align*}
Plugging in the expression for $x_{t-L}$ from \eqref{eq.x_{t-L}},
\begin{align*}
\begin{aligned}
y_{t} &= C A^L \Ocal_L^{\dagger} \left( Y_{t}
- \tilde{\Tcal}_L E_{t} - \left( I_L \otimes D \right) F_{t}\right)\\
&\qquad + \sum_{j=0}^{{L-1}} M_j \psi_1(u_{t-1-j})+ D \psi_2 (u_{t}).
\end{aligned}
\end{align*}
The matrix $Q:=C A^L \Ocal_L^{\dagger} \in \Rset^{p \times pL}$  can be written in block form as $Q=[Q_{L-1}, \cdots, Q_{0}]$ with $Q_i \in \Rset^{p \times p}$ for $i = 0,\cdots,{L-1}$. Setting $A_k:= - Q_{k}$ for $k = 1,\cdots,L-1$ and moving them to the LHS of the above equation, we have
\begin{align*}
\begin{aligned}
y_{t} + \sum_{k=1}^{L} A_k y_{t-k}
=& \sum_{j=0}^{{L-1}} M_j \psi_1(u_{t-1-j})
- Q \tilde{\Tcal}_L E_{t} \\
&- Q \left( I_L \otimes D \right) F_{t} + D \psi_2 (u_{t}).
\end{aligned}
\end{align*}
With $M = [M_{{L-1}}, \cdots M_0] \in \Rset^{p \times mL}$, we can write the RHS as a function of on $u_{t-L:t}$ as
\begin{align*}
\begin{aligned}
&y_{t} + \sum_{k=1}^{L} A_k y_{t-k} \\
&= \underbrace{\begin{bmatrix} M- Q \tilde{\Tcal}_L & - Q \left( I_L \otimes D \right) & D \end{bmatrix}}_{=: S}
\underbrace{\begin{bmatrix} E_{t} \\ F_{t} \\ \psi_2 (u_{t})\end{bmatrix}}_{=: \psi(u_{t-L:t})} \\
&=: g(u_{t-L:t}),
\end{aligned}
\end{align*}
where $S \in \Rset^{p \times q(2L+1)}$, $\psi(u_{t-L:t}) \in \Rset^{q(2L+1)}$, and $g$ defined above is a  mapping from $\Usf$ into $\Ycal = \Rset^p$.
Assume now that $\psi_1,\psi_2$ are elements of some $\Rset^q$-valued RKHS $\Hcal$ on $\Ucal$. Thus, by definition, the evaluation maps $\psi_1 \mapsto \psi_1(u)$ and $\psi_2 \mapsto \psi_2(u)$ are bounded for each $u \in \Ucal$. Consider the direct-sum Hilbert space $\Kcal := \Hcal^{\oplus(2L+1)}$ with inner product $\langle h,h'\rangle_{\Hcal^{\oplus(2L+1)}}=\sum_{j=0}^{2L} \langle h_j,h'_j\rangle_{\Hcal}$ for $h=h_0 \oplus \cdots \oplus h_{2L}$. The function $g$ defined above is an element of the linear space $\Fcal$ of functions $f : \Usf \to \Ycal $ of the form
\begin{align*}
f(u_{0:L})  = S_0 h_0(u_0) + \sum^{L}_{j=1} S_j h_j(u_j) + \sum^{2L}_{j=L+1} S_{j} h_j(u_{j-L}),
\end{align*}
where $h = h_0 \oplus \dots \oplus h_{2L}$ ranges over $\Kcal$ and where $S_0, S_1, \dots, S_{2L}$ are linear operators from $\Rset^q$ into $\Ycal = \Rset^p$. For each $u_{0:L}$, the evaluation map $f \mapsto f(u_{0:L})$ is continuous on $\Fcal$. Thus, $g$ belongs to some $\Ycal$-valued RKHS on $\Usf$.

\section{Behavior Representer Theorem Via Minimum Norm Interpolation}
\label{sec.interpolate}

Let $(u_t^\data,y_t^\data)_{t=0}^{T+L-1}$ be a finite input-output trajectory generated by an unknown system of the form \eqref{eq.NL.AR}. In other words, there exists an unknown $(L+1)$-tuple $(A_1,\dots,A_L,g)$, such that
\begin{align*}
\begin{bmatrix}
u^\data_{0:T+L-1} \\ y^\data_{0:T+L-1} 
\end{bmatrix} \in \Bcal(A_1,\dots,A_L,g)|_{T+L-1}.
\end{align*}
The problem of data-driven behavioral modeling is to reconstruct the unknown length-$L$ behavior segment $\Bcal(A_1,\dots,A_L,g)|_{L}$ from the data without explicitly estimating $A_1,\dots,A_L,g$.

Let $f$ be the regression representation of $(A_1,\dots,A_L,g)$ as in \eqref{eq.NL.AR.f}. Using the definitions of $z_t$ in \eqref{eq:regression_vectors} and $y_{t^+} := y_{t+L}$, we can represent the data $(u^\data_t,y^\data_t)^{T+L-1}_{t=0}$ equivalently by $(z^\data_t,y^\data_{t^+})^{T-1}_{t=0}$, such that $f(z^\data_t) = y_{t^+}$ holds for all $t = 0,\dots,T-1$. In other words, for each $t = 0,\dots,T-1$, $(\kappa_\Zcal(\cdot,z^\data_t)^*,y^\data_{t^+})$ is a valid i/o pair for \eqref{eq.NL.AR.f}. In view of the equivalence \eqref{eq.behavior.equivalence}, the question we would like to answer is whether we can reconstruct the set of \textit{all} valid i/o pairs of the unknown $f$ from the given data.

The following result is easy to establish:
\begin{theorem}
For any collection of real coefficients $c_0,\dots,c_{T-1}$, 
$\left(\sum_{j=0}^{T-1}c_j \kappa_\Zcal(\cdot,z_j^\data)^*,\sum_{j=0}^{T-1}c_j y_{j^+}^\data \right)$ is a valid i/o pair for the model \eqref{eq.NL.AR.f}. 
\label{theorem.NL.<=}
\end{theorem}

\noindent Theorem~\ref{theorem.NL.<=} shows that any linear combination of $\big(\kappa_\Zcal(\cdot,z^\data_t)^*, y^\data_{t^+ }\big)^{T-1}_{t=0}$ is a valid i/o pair of \eqref{eq.NL.AR.f}. We now analyze the reverse direction by relating it to minimum-norm interpolation in a vector-valued RKHS \cite{micchelli2005learning}. Let finite input-output data $\{(z_i,y_{i })\}^{N-1}_{i=0} \subset \Zcal \times \Ycal$ be given and consider the following problem:
\begin{align}
\begin{aligned}
\text{minimize } & \vnorm{f}_{\Hcal_\Zcal}^2 \\
\text{subject to } & \ f\left(z_j\right) = y_{j }, \quad  j =0,\cdots,{N-1}.
\end{aligned}
\label{eq.MNIP.v}
\end{align}
Define the \textit{sampling operator} $S_N : \Hcal_\Zcal \to \Ycal^N$  by $S_N f = \left(f(z_0),\cdots, f(z_{N-1})\right)$. By \cite[Theorem 3]{micchelli2005learning}, if $(y_0,\dots,y_{N-1}) \in \range(S_N)$,
then the minimum norm interpolation problem in \eqref{eq.MNIP.v} has a unique solution given by
\begin{align}
f_N = \sum_{j=0}^{N-1} \kappa_\Zcal(\cdot,z_j) v_j,
\label{eq.f.star}
\end{align}
where $(v_0,\dots,v_{N-1}) \in \Ycal^N$ solves the system of equations
\begin{align}\label{eq.normal}
\sum_{l=0}^{N-1} \kappa_\Zcal(z_j,z_l) v_l = y_{j }, \qquad j = 0,\dots,N-1.
\end{align}
We can express this more succinctly as follows.

Since $\Ycal = \Rset^p$, for each $z \in \Zcal$ we can view $\kappa_\Zcal(\cdot,z)$ as a mapping from $\Rset^p$ into $\Hcal_\Zcal$, i.e., for each $v \in \Rset^p$, $\kappa_\Zcal(\cdot,z) v \in \Hcal_\Zcal$. Define the mapping $\Phi_N:\Rset^{pN} \to \Hcal_\Zcal$ as 
\begin{align*}
\Phi_N \bar{v} = \sum_{j=0}^{N-1} \kappa_\Zcal(\cdot,z_j) v_j,
\end{align*}
where $\bar{v} = (v_0,\cdots,v_{N-1}) \in  \Rset^{pN}$. In particular, $f_N = \Phi_N\bar{v}$, where $\bar{v}$ is the solution of \eqref{eq.normal}. Next, define the block kernel matrix $K_N := \Phi_N^* \Phi_N \in \Rset^{pN \times pN}$ whose blocks are given by $[K_N]_{ij} = \kappa_\Zcal(z_i,z_j)$ for $i,j = 0,\cdots,N-1$, as well as  the block row vector $k_N(z):=\kappa_\Zcal(\cdot,z)^*\Phi_N \in \Rset^{p\times pN}$ with blocks $[k_N(z)]_j = \kappa_\Zcal(z,z_j)$ for $j = 0,\cdots,N-1$. Using the reproducing property and the above definitions, we can write
\begin{align}
f_N(z) &= \kappa_\Zcal(\cdot,z)^* f_N \nonumber\\
&= \kappa_\Zcal(\cdot,z)^* \Phi_N K^\dagger_N y_{N } \nonumber\\
&= k_N(z) K^\dagger_N Y_{N}, \label{eq:kKY_formula}
\end{align}
where $Y_{N} := [y^\trn_{0 },\dots,y^\trn_{N-1 }]^\trn$. Finally, for each $z$ define $\Sigma_N(z) \in \Rset^{p \times p}$ as
\begin{align}
\Sigma_N(z) := \kappa_\Zcal(z,z) - k_N(z)K_N^\dagger {k}_{N}(z)^*.
\label{def.Sigma}
\end{align}

We now present two key lemmas. The first one is a structural characterization of $\Sigma_N$:
\begin{lemma}
For each $z$, the operator $\Sigma_N(z)$ is symmetric and positive semi-definite, and 
$\Sigma_N(z) = 0$ iff ${\rm range}\left(\kappa_\Zcal(\cdot,z)\right) \subseteq {{\rm range}(\Phi_N)}$.
\label{lemma.Sigma.t}
\end{lemma}
\noindent The second one is an extension of a result of Liang and Recht \cite[Lemma~1]{liang2023interpolating} to the minimum-norm interpolation problem in the vector-valued RKHS $\Hcal_\Zcal$:
\begin{lemma} Suppose that the problem \eqref{eq.MNIP.v} admits unique solutions $f_{N}$ and $f_{N+1}$ given the respective data $\{(z_i,y_{i })\}^{N-1}_{i=0}$ and $\{(z_i,y_{i })\}^{N-1}_{i=0} \cup \{ (z_{N},y_{N }) \}$. Then the following holds:
\begin{enumerate}
\item If $\Sigma_{N}(z_N) = 0$, then $y_{N } - f_N\left(z_N\right) = 0$.
\item If $\Sigma_{N}(z_N) \succ 0$, we have
\begin{align}
\begin{aligned}
& \vnorm{f_{N+1}}_{\Hcal_\Zcal}^2 - \vnorm{f_{N}}_{\Hcal_\Zcal}^2 \\
& \qquad = \vnorm{\Sigma_{N}(z_N)^{-1/2} \left(y_{N } - f_{N}\left(z_N\right) \right)}_{\Ycal}^2.
\end{aligned}\label{eq:prediction_error}
\end{align}
\end{enumerate}
\label{lemma.liang.vRKHS}
\end{lemma}

\noindent Lemma~\ref{lemma.liang.vRKHS} characterizes the error incurred when we use $\hat{y}_{N } = f_N(z_N)$ as a \textit{predictor} of $y_{N } = f_{N+1}(z_N)$,  where $f_N$ is the solution to the minimum norm interpolation problem on $\{(z_i,y_{i })\}^{N-1}_{i=0}$. In particular, the prediction is exact when $\Sigma_N(z_N) = 0$. If $\Sigma_N(z_N)$ is nonzero but still positive definite, the identity \eqref{eq:prediction_error} expresses the squared norm of the weighted prediction error $\Sigma_N(z_N)^{-1/2}(y_{N } - \hat{y}_{N })$ in terms of the norms of $f_{N+1}$ and $f_N$.

In a scalar-valued RKHS, $\Sigma_N(z)$ is equal to
\begin{align}\label{eq.SigmaN_scalar}
\begin{split}
&{\rm dist}^2(\kappa_\Zcal(\cdot,z),{\rm span} (\kappa_\Zcal(\cdot,z_0),\cdots , \kappa_\Zcal(\cdot,z_{N-1}) )) \\
&= \min_{h \in {\rm span} (\kappa_\Zcal(\cdot,z_0),\cdots , \kappa_\Zcal(\cdot,z_{N-1}) )} \| \kappa_\Zcal(\cdot,z) - h \|_{\Hcal_\Zcal}^2,
\end{split}
\end{align}
which is the squared distance from $\kappa_\Zcal(\cdot,z)$ to the linear span of kernel functions centered at the observed samples \cite{liang2023interpolating}. While Lemma~\ref{lemma.liang.vRKHS} applies to \textit{any} vector-valued RKHS of $\Ycal$-valued functions on $\Hcal_\Zcal$, a natural choice of the operator-valued kernel could be $\kappa_\Zcal(z_1,z_2) = \kappa_\Zcal^\s (z_1,z_2) I_{\Ycal}$, where $\kappa_\Zcal^\s$ is a scalar-valued kernel function. Let ${K}_N^\s = [\kappa_\Zcal^s(z_i,z_j)]^{N-1}_{i,j=0}\in \Rset^{ N \times N}$ be the Gram matrix defined by the scalar-valued kernel $\kappa_\Zcal^\s $ on the points $z_0,\dots,z_{N-1}$, let $k^\s_N(z) \in \Rset^{1\times N} $ be the row vector with $[k_N^\s(z)]_j = \kappa_\Zcal^\s (z,z_j)$,  ${K}_N = {K}^\s_N \otimes I_{\Ycal}$, and $K_N(z_N) = k^\s_{N-1}(z) \otimes I_{\Ycal}$. 
Finally, define
\begin{align*}
s_N := {\rm dist} \left\{ {\rm span} \left(\kappa_\Zcal^\s(\cdot,z_0),\cdots , \kappa_\Zcal^\s(\cdot,z_{N-1}) \right), \kappa_\Zcal^\s(\cdot,z_{N})\right\},
\end{align*}
where the distance is computed in the scalar-valued RKHS induced by $\kappa_\Zcal^\s$, cf.~\eqref{eq.SigmaN_scalar}. Then, using the properties of the Kronecker product $\otimes$, we can compute $\Sigma_N(z_N)$ as follows:
\begin{align*}
&\Sigma_N(z_N) \\
&=\kappa_\Zcal(z_N,z_N) - K_N(z_N)K_N^\dagger K_N(z_N)^*\\
&=\kappa_\Zcal^\s(z_N,z_N) I_{\Ycal}\\
&\,\,- \left( k^\s_{N-1}(z_N) \otimes I_{\Ycal} \right)^* \left({K}^\s_{N-1} \otimes I_{\Ycal} \right)^\dagger 
\left(k^\s_{N-1}(z_N) \otimes I_{\Ycal} \right) \\
&= \left(\kappa_\Zcal^\s(z_N,z_N) -k^\s_{N-1}(z_N)^\trn ({{K}^\s_{N-1}})^\dagger k^\s_{N-1}(z_N) \right)  I_{\Ycal}\\
&= s_{N}^2 I_{\Ycal}.
\end{align*}
Hence, the equality in Lemma \ref{lemma.liang.vRKHS} reduces to
\begin{align*}
\begin{aligned}
s_N^2\left(\vnorm{f_{N+1}}_{\Hcal_\Zcal}^2 - \vnorm{f_{N}}_{\Hcal_\Zcal}^2\right) 
= \vnorm{y_{N } - f_{N}\left(z_N\right)}_{\Ycal}^2.
\end{aligned}
\end{align*}
This indicates that, given a new $(z_N,y_{N})$, if $s_N=0$, we have $\vnorm{y_{N} - f_N(z_N)}_2^2 = 0$. Given Lemmas \ref{lemma.Sigma.t} and \ref{lemma.liang.vRKHS}, the following is immediate: 
\begin{theorem}[Behavior Representer Theorem]\label{thm.behavior.rep} Let $\left(z^\data_t,y^\data_{t^+} \right)_{t=0}^{T-1}$ be a length-$T$ trajectory of regression and output vectors for an unknown $f_\star \in \Hcal_\Zcal$, i.e.,
\begin{align*}
y^\data_{t^+} = f_\star(z^\data_t) \qquad \text{for each } t = 0,\dots,T-1.
\end{align*}
Let $[u_{0:L}^\trn, y_{0:L}^\trn]^\trn$ be an element of $\Bcal(f_\star)|_L$, and let $(z_0, y_{0^+})$ be computed according to \eqref{eq.z0_y0plus}. Then the following holds:
\begin{enumerate}
    \item If $\Sigma_T(z_0) = 0$, then  $y_{0^+} = k_T(z_0) K_T^\dagger Y_{T}$, where $K_T$ and $Y_T$ are computed from the data according to \eqref{eq:kKY_formula} and $Y_{T}:=[(y^\data_{0^+})^\trn,\cdots,(y^\data_{(T-1)^+})^\trn]^\trn$.
    \item If $\Sigma_T(z_0) \succ 0$, then 
    \begin{align}\label{eq.rec_error}
    \begin{split}
& \| \Sigma_T(z_0)^{-1/2}(y_{0^+} - k_T(z_0) K_T^\dagger Y_{T}) \|^2_\Ycal \\
&\qquad \qquad  = \| f_{T+1} \|^2_{\Hcal_\Zcal} - \| f_T \|^2_{\Hcal_\Zcal} \\
& \qquad \qquad \le \| f_\star \|^2_{\Hcal_\Zcal} - \| f_T \|^2_{\Hcal_\Zcal},
\end{split}
    \end{align}
    where $f_T$ (respectively, $f_{T+1})$ is the minimum-norm interpolator of $\{(z^\data_t,y^\data_{t^+})\}^{T-1}_{t=0}$ (respectively, of $\{(z^\data_t,y^\data_{t^+})\}^{T-1}_{t=0} \cup \{(z_0, y_{0^+})\}$.
\end{enumerate}
\label{thm.NL.interpolate}
\end{theorem}
\noindent The condition $\Sigma_T(z_0) = 0$ describes the setting when exact reconstruction is possible. By Lemma~\ref{lemma.Sigma.t}, this will be the case for all pairs $(z,y_+=f_\star(z))$ for which ${\rm range}(\kappa_\Zcal(\cdot,z)) \subseteq {\rm range}(\Phi_T) \}$. If $z_0$ does not satisfy this condition but $\Sigma_T(z_0) \succ 0$, we can quantify the reconstruction error using \eqref{eq.rec_error}. For LTI systems, a sufficient condition for the existence of unique minimum-norm interpolating solutions is 
	\begin{align*}
		\sum^{T-1}_{t=0} z^\data_t (z^\data_t)^\trn \succ 0,
	\end{align*}
which is the classical persistence of excitation condition \cite{moore1983PE,Green1986PE}. Exact reconstruction is possible when $z_0 \in {\rm span}(z^\data_0,\dots,z^\data_{T-1})$. In the nonlinear setting, the condition $\Sigma_T(z_0) = 0$ plays an analogous role via Lemma \ref{lemma.Sigma.t}.

It is useful to compare Theorem~\ref{thm.behavior.rep} with the fundamental lemma for LTI systems (cf.~Section~\ref{sec.preliminary}). The latter characterizes behaviors through the image of the Hankel matrix built from observed input-output trajectories. Theorem~\ref{thm.behavior.rep} of this section is conceptually analogous, but is phrased in terms of a different data representation based on regression vectors, which are formed from inputs and past outputs. 

\section{Subspace Identification in the Vector-Valued RKHS Setting}
\label{sec.subspace}

We now revisit systems that arise from state-space representations, as in Section~\ref{sec.NL.ss}. We will focus on a particular class of such systems, namely ones that can be represented as
\begin{align}
\begin{aligned}
x_{t+1} &= A x_{t} + B \phi\left(u_{t}\right),\\
y_{t} &= C x_{t} + D \phi\left(u_{t}\right),
\end{aligned}
\label{eq.NL.state-space.phi}
\end{align}
where $\phi: \Ucal \to \Rset^q$ is a mapping from the input space $\Ucal = \Rset^m$ into $\Rset^q$. 

Given $(A,B,C,D)$, we construct the $L$-step controllability matrix $\Ccal_L$, the $L$-step observability matrix $\Ocal_L$, the reversed $L$-step controllability matrix $\Delta_L$, and the $L$-step modified Toeplitz matrix $\tilde{\Tcal}_L$ exactly as in \eqref{def.observablity}. The $L$-step Toeplitz matrix is given by
\begin{align*}
\Tcal_L := \tilde{\Tcal}_L + I_L \otimes D.
\end{align*}
Let  $\left\{(u^\data_t,y^\data_t\right)\}_{t=0}^{T-1}$ denote input/output data of length $T$ collected from measurements of \eqref{eq.NL.state-space.phi} starting from some initial condition $x_0 \in \Rset^n$. In the LTI case (i.e., when $q = m$ and $\phi$ is the identity map),  subspace identification methods  \cite{van2012subspace} allow one to reconstruct,  under certain regularity conditions, the state trajectory $x_L,\dots,x_{T-L}$ and the observability matrix $\Ocal_L$ directly from the input/output data without knowledge of the system matrices $A,B,C,D$. 
 
In this section, we show that these methods can be extended to the set-up of \eqref{eq.NL.state-space.phi} when $\phi$ is an element of a suitable vector-valued RKHS on  $\Ucal$. Moreover, we obtain a result in the spirit of the fundamental lemma for LTI systems, namely that the set of all valid length-$L$ input/output trajectories of \eqref{eq.NL.state-space.phi} can be reconstructed directly from the data $\left\{(u^\data_t,y^\data_t\right)\}_{t=0}^{T-1}$ without an intermediate system identification step.

\subsection{The construction of a vector-valued RKHS}
\label{sec.state-space.phi}

There are various ways of instantiating the state-space model \eqref{eq.NL.state-space.phi} in the vector-valued RKHS framework presented in Section~\ref{sec.examples}, i.e., choosing a suitable $\Rset^q$-valued RKHS $\Hcal_\kappa$ on the input space $\Ucal$ with reproducing kernel $\kappa$ so that $\phi \in \Hcal_\kappa$. Perhaps the simplest one is to define the operator-valued map $\kappa : \Ucal \times \Ucal \to \Lcal(\Rset^q)$ by
\begin{align*}
\kappa(u,u') := \phi(u) \otimes \phi(u'),
\end{align*}
which is readily seen to be of the positive type since, for any $n$, any $\alpha_1,\dots,\alpha_n \in \Rset$, any $u_1,\dots,u_n \in \Ucal$, and any $v \in \Rset^q$,
\begin{align*}
\begin{aligned}
& \sum^n_{i,j=1}\alpha_i \alpha_j \langle \kappa(u_i,u_j)v, v \rangle_{\Rset^q} \\
&= \sum^n_{i,j=1}\alpha_i \alpha_j \langle \phi(u_i),v \rangle_{\Rset^q} \langle \phi(u_j), v\rangle_{\Rset^q} \\
&= \left( \sum^n_{i=1} \alpha_i \langle \phi(u_i), v \rangle_{\Rset^q}\right)^2 \\
&\ge 0.
\end{aligned}
\end{align*}
By \cite[Prop.~2.3]{carmeli_2006}, there is a unique $\Rset^q$-valued RKHS $\Hcal_\kappa$ of functions on $\Ucal$ with reproducing kernel $\kappa$. The kernel $\kappa$ has the convenient property 
\begin{align}
\left \langle \phi(u), \phi(u') \right \rangle_{\Rset^q}
=\Tr\left(\kappa(u,{u'})\right).
\label{eq.phi.innerproduct}
\end{align}
\subsection{Subspace identification using RKHS methods}

We now have all the ingredients in place for putting together a subspace identification framework analogous to the one for LTI systems \cite{van2012subspace}.

Let the input/output data  $\left\{(u^\data_t,y^\data_t\right)\}_{t=0}^{T-1}$ be given. 
Introduce the Hankel-type block matrices
\begin{align}
\begin{aligned}
\begin{bmatrix} {Y_{\past}} \\ \midrule {Y_{\future}} \end{bmatrix}
:=
\begin{bmatrix}
 y^\data_0 & 
\cdots  &  y^\data_{T-2L}\\
\vdots &  
&\vdots \\
 y^\data_{L-1} & 
\cdots &  y^\data_{T-L-1}\\
\midrule
 y^\data_{L}& 
\cdots &  y^\data_{T-L}\\
\vdots &  
&\vdots\\
 y^\data_{2L-1} & 
\cdots &  y^\data_{T-1}
\end{bmatrix}.
\end{aligned}
\label{eq.def.Y}
\end{align}
and
\begin{align}
\begin{aligned}
\begin{bmatrix} U_{\past}^\phi \\ \midrule U_{\future}^\phi \end{bmatrix}
:=\begin{bmatrix}
\phi \left(u^\data_0\right) & 
\cdots & \phi \left(u^\data_{T-2L}\right) \\
\vdots &  
&\vdots & \\
\phi \left(u^\data_{L-1}\right) & 
\cdots & \phi \left(u^\data_{T-L-1}\right) \\
\midrule
\phi \left(u^\data_{L}\right) & 
\cdots & \phi \left(u^\data_{T-L}\right) \\
\vdots &  
&\vdots & \\
\phi \left(u^\data_{2L-1}\right) & 
\cdots & \phi \left(u^\data_{T-1}\right)
\end{bmatrix},
\end{aligned}
\label{eq.H.2L}
\end{align}
where $\past$ and $\future$ designates a partition into ``past'' and ``future'' data.
Let $H_{\past}$ and $H_{\future}$ denote, respectively, the concatenations of $U_{\past}$ and $Y_{\past}$ and of $U_{\future}$ and $Y_{\future}$:
\begin{align*}
H_{\past}:=\begin{bmatrix} U_{\past} \\ Y_{\past} \end{bmatrix}, \quad H_{\future}:=\begin{bmatrix} U_{\future} \\ Y_{\future} \end{bmatrix}.
\end{align*}
Similar to the linear case, let $\Pi$ denote the oblique projection of the rowspace of $Y_{\future}$ onto the rowspace of $H_{\past}$ along the rowspace of $U_{\future}$:
\begin{align*}
\Pi:= {Y_{\future}}  \underset{U_{\future}^\phi}{/} H_{p},
\end{align*}
which satisfies $\rank(\Pi)\leq T-2L+1$. Let the SVD of $\Pi$ be given by 
\begin{align*}
\Pi= \begin{bmatrix}
U_1 & U_2 \end{bmatrix} \begin{bmatrix}
\Sigma_1 & 0 \\ 0 & 0 \end{bmatrix} \begin{bmatrix}
V_1^{\trn} \\ V_2^{\trn} \end{bmatrix} = U_1 \Sigma_1 V_1^{\trn}.
\end{align*}
Following \cite{van2012subspace}, let $X_\future := \begin{bmatrix} x_L & x_{L+1} & \dots & x_{T-L} \end{bmatrix}$, where $x_t$ is the state trajectory of \eqref{eq.NL.state-space.phi} determined by the initial state $x_0$ and the inputs $u^\data_t$. In the LTI case, $X_\future$ can be recovered, up to a similarity transformation, via the formula $X_{\future}= \Sigma_1^{1/2} V_1^\trn$ (see, e.g., \cite[Theorem 2, Chapter 2]{van2012subspace}). In fact, the same result holds in the present nonlinear case (the idea is to view $v_k := \phi(u_k)$ as an input to an LTI system given by $(A,B,C,D)$). Given the input/output data, let
\begin{align}
U^\phi
:=\begin{bmatrix}
\phi \left(u^\data_{0}\right)  &\cdots  & \phi \left(u^\data_{T-2L-n}\right)\\
\vdots &   &\vdots  \\
\phi \left(u^\data_{2L-1}\right) & \cdots & \phi \left( u^\data_{T-n}\right)\\
\midrule
\phi \left(u^\data_{2L}\right) & \cdots & \phi \left(u^\data_{T-n}\right)\\
\vdots &   &\vdots  \\
\phi \left(u^\data_{2L-1+n}\right) & \cdots & \phi \left(u^\data_{T-1}\right)
\end{bmatrix},
\label{eq.H.2L+n}
\end{align}
and form the input Gram matrix of depth $2L+n$ as $K^u:=(U^\phi)^{\trn}U^\phi$ with entries 
$[K^u]_{ij} = \sum_{k=0}^{2L+n-1}\kappa\left( u^\data_{i+k},u^\data_{j+k}\right)$.
The following theorem is a straightforward adaptation of  subspace identification results for LTI systems:
\begin{theorem}
Let $\left\{u^\data_i,y^\data_i\right\}_{i=0}^{T-1}$ be a sequence of length $T$ generated by the system \eqref{eq.NL.state-space.phi}, where $(A,B)$ is controllable and $(A,C)$ is observable.  Suppose that the input sequence is such that 
$\rank(K^u) = (2L+n) q$.
Then, $\Pi= \Ocal_L X_{\future}$ and $X_{\future} = \Sigma^{1/2}_1 V_1^\trn$.
\label{thm.nonlinear.realization}
\end{theorem}
\noindent The process of computing the oblique projection and constructing the state vector can be carried out using Gram matrices computed from pairwise kernel evaluations $\kappa$, without explicitly using $\phi$. Specifically, define the Gram matrices $K_\past^u= \left({U_{\past}^\phi}\right)^{\trn}{U_{\past}^\phi}$, $K_\future^u= \left({U_{\future}^\phi}\right)^{\trn}{U_{\future}^\phi}$, and $K_\past^y= {Y_{\past}}^{\trn}{Y_{\past}}$, whose entries are
\begin{align*}
\begin{aligned}
[K_\past^u]_{ij} =& \sum_{k=0}^{L-1} \left \langle \phi(u^\data_{i+k}), \phi(u^\data_{j+k}) \right \rangle_{\Rset^M}\\
\stackrel{(a)}{=} & \sum_{k=0}^{L-1} \Tr \left( \kappa\left(u^\data_{i+k}, u^\data_{j+k}\right) \right),\\
[K_\future^u]_{ij} =& \sum_{k=0}^{L-1} \left \langle \phi(u^\data_{i+L+k}), \phi(u^\data_{j+L+k}) \right \rangle_{\Rset^M}\\
\stackrel{(b)}{=} & \sum_{k=0}^{L-1} \Tr \left( \kappa\left(u^\data_{i+L+k},u^\data_{j+L+k} \right)\right),\\
[K_\past^y]_{ij}=&\sum_{k=0}^{L-1} {y^\data_{i+k}}^\trn y^\data_{j+k},
\end{aligned}
\end{align*}
respectively, where (a) and (b) follow from \eqref{eq.phi.innerproduct}.

The oblique projection can be computed as
\begin{align}
\begin{aligned}
\Pi=&{Y_{\future}} \underset{{U_{\future}^\phi}}{/} H_{p}  \\ 
=& {Y_{\future}} \begin{bmatrix} {H_{\past}}^{\trn} & ({U_{\future}^\phi})^{\trn} \end{bmatrix} 
\left( \begin{bmatrix}  {H_{\past}} \\ {{U_{\future}^\phi}} \end{bmatrix} \begin{bmatrix}  {H_{\past}}^{\trn} & ({U_{\future}^\phi})^{\trn} \end{bmatrix} \right)^\dagger \begin{bmatrix} {H_{\past}} \\ {0}
 \end{bmatrix} \\
=& {Y_{\future}} \left(\begin{bmatrix} {H_{\past}}^{\trn} & ({U_{\future}^\phi})^{\trn} \end{bmatrix} 
\begin{bmatrix}  {H_{\past}} \\ {{U_{\future}^\phi}} \end{bmatrix} \right)^\dagger \begin{bmatrix}  {H_{\past}}^{\trn} & ({U_{\future}^\phi})^{\trn} \end{bmatrix} 
\begin{bmatrix} {H_{\past}} \\ {0} \end{bmatrix} \\
=& {Y_{\future}} \left(K_{\past}^u + K_{\past}^y + K_{\future}^u \right)^{\dagger}
\left(K_{\past}^u + K_{\past}^y \right), \\
\end{aligned}
\label{eq.nonlinear.oblique}
\end{align}
where the second-to-last line follows from the identity $x (x^{\trn} x)^{\dagger} = (x x^{\trn})^\dagger x$.

To construct the state vector, instead of directly performing SVD on $\Pi$, we construct $U$, $\Sigma$, and $V$ such that $\Pi = U \Sigma V^{\trn}$ via eigendecomposition of $\Pi^{\trn} \Pi$ and $\Pi \Pi^{\trn}$. Denote $\overline{K}_{\past,\future}:= K_{\past}^u + K_{\past}^y + K_{\future}^u$ and $\overline{K}_{\past} := K_{\past}^u + K_{\past}^y$ and notice that we have
\begin{align*}
\begin{aligned}
\Pi^{\trn} \Pi 
=& \overline{K}_{\past}^{\trn} \left(\overline{K}_{\past,\future}^{\dagger}\right)^{\trn} \left({{Y_{\future}}}^{\trn} {{Y_{\future}}}\right)\overline{K}_{\past,\future}^{\dagger} \overline{K}_{\past} \\
=& \overline{K}_{\past}^{\trn} \left(\overline{K}_{\past,\future}^{\dagger}\right)^{\trn} K_{\future}^y \overline{K}_{\past,\future}^{\dagger} \overline{K}_{\past} \\
=& \overline{K}_{\past}^{\trn} \overline{K}_{\past,\future}^{\dagger} K_{\future}^y \overline{K}_{\past,\future}^{\dagger} \overline{K}_{\past}.
\end{aligned}
\end{align*}
Then, we can obtain $\left(\sigma_i^{1/2}, v_i\right)$ as the eigenvalues and right eigenvectors of $\Pi^{\trn} \Pi$. Likewise, we have
\begin{align*}
\begin{aligned}
\Pi \Pi^{\trn}
=& {Y_{\future}} \overline{K}_{\past,\future}^{\dagger}
\overline{K}_{\past} \overline{K}_{\past}^{\trn} \overline{K}_{\past,\future}^{\dagger} {{Y_{\future}}}^{\trn}.
\end{aligned}
\end{align*}
Denote $\Gamma:= \overline{K}_{\past,\future}^{\dagger} \overline{K}_{\past} \overline{K}_{\past}^{\trn} \overline{K}_{\past,\future}^{\dagger}$. Then, we have
\begin{align*}
{Y_{\future}} \Gamma {{Y_{\future}}}^{\trn} \left({Y_{\future}} z_i\right) = \sigma^{1/2}_i \left({Y_{\future}} z_i\right)
\Leftrightarrow 
\Gamma \underbrace{{{Y_{\future}}}^{\trn} {Y_{\future}}}_{=: K_{\future}^y} \xi_i= \sigma_i^{1/2}  \xi_i.
\end{align*}
That is, $\Pi \Pi^{\trn}={{Y_{\future}}} \Gamma {{Y_{\future}}}^{\trn}$ has an eigenvector $u_i = {Y_{\future}} \xi_i$ associated with eigenvalue $\sigma^{1/2}_i$ iff $\xi_i$ is an eigenvector of $\Gamma K_{\future}^y$ corresponding to the same eigenvalue.
With $U=[u_1,\cdots,u_{n}]$, $\Sigma = \text{diag}(\sigma_1,\cdots, \sigma_n)$, and $V=[v_1,\cdots, v_n]$, we can define the extended observability matrix $\Ocal_L$ and states $X_{f}$ (up to a similarity transform) as
\begin{align*}
\Ocal_L = U \Sigma^{1/2}, \quad
X_{\future} = V \Sigma^{1/2}.
\end{align*}
The following corollary is immediate from the above process.
\begin{corollary}[Construction of state vectors]
Suppose that $(A,B)$ is controllable and $(A,C)$ is observable, i.e., $\rank(\Ocal_L) = \rank(\Ccal_L) = n$.  Let input-output data $\{(u^\data_t,y^\data_t)\}_{t=0}^{T-1}$ of length $T$ be given.
Suppose the input sequence is such that 
$\rank(K_{2L+n}^u) = (2L+n) q$. Let $(\Sigma^{1/2},V)$ be the eigenpair of $\overline{K}_{\past}^{\trn} \left({\overline{K}_{\past,\future}}^{-1}\right)^{\trn} K_{\future}^y {\overline{K}_{\past,\future}}^{-1} \overline{K}_{\past}$, and $(\Sigma^{1/2},\Xi)$ be the eigenpair of ${\overline{K}_{\past,\future}}^{-1} \overline{K}_{\past} \overline{K}_{\past}^{\trn} \left({\overline{K}_{\past,\future}}^{-1}\right)^{\trn} K_{\future}^y$. 
Define $U ={Y_{\future}} \Xi$.
Then the extended observability matrix is ${\Ocal}_L = U \Sigma^{1/2}$, and the state sequence is $X_{f}= V\Sigma^{1/2} $ (both up to a similarity transformation).
\label{corollary.nonlinear.realization.2}
\end{corollary}

The next result explicitly connects subspace identification to an RKHS version of the fundamental lemma for nonlinear systems admitting state-space realization \eqref{eq.NL.state-space.phi}.

\begin{theorem}  Consider the state-space model \eqref{eq.NL.state-space.phi}, where $(A,B)$ is controllable and $(A,C)$ is observable. Let input-output data $\{(u^\data_t,y^\data_t)\}_{t=0}^{T-1}$ of length $T$ be given, and suppose the input Gram matrix is such that
$\rank(K_{2L+n}^u )= (2L+n) q$ and that $L$ is chosen so that $\Ocal_L$ has full column rank. Then a length-$2L$ sequence 
$\begin{bmatrix} u_{0:2L-1} \\ y_{0:2L-1} \end{bmatrix}$ is a valid input/output trajectory of the system \eqref{eq.NL.state-space.phi} if and only if
there exists $\xi \in \Rset^{T-2L+1}$ such that
\begin{align*}
\begin{bmatrix} k_{\past}^u  \\ k_{\past}^y \end{bmatrix} 
= \begin{bmatrix} K_{\past}^u\\ K_{\past}^y \end{bmatrix} \xi
\text{ and }
\begin{bmatrix} k_{\future}^u\\k_{\future}^y \end{bmatrix} 
=\begin{bmatrix} K_{\future}^u \\ K_{\future}^y \end{bmatrix} \xi,
\end{align*}
where, for $\Phi_{0:2L-1} := [\phi(u_0)^\trn, \dots, \phi(u_{2L-1})^\trn]^\trn$, the vectors $k_{\past}^u = (U_{\past}^\phi)^{\trn} \Phi_{0:2L-1}$ and $k_{\future}^u = (U_{\future}^\phi)^{\trn} \Phi_{0:2L-1}$ have entries given by $[k_{\past}^u]_j = \sum_{r=0}^{L-1} \kappa(u^\data_{j+r}, u_r)$, $[k_{\future}^u]_j = \sum_{r=0}^{L-1} \kappa(u^\data_{j+L+r}, u_{L+r})$; and $k_{\past}^y = Y_\past^{\trn} {y}_{0:L-1}$, $k_{\future}^y = Y_\future^{\trn} {y} _{L:2L-1}$ are the Euclidean inner products of outputs.
\label{thm.NL.FL}
\end{theorem}

In the context of the above theorem, we can view $(K_{\past}^u, K_{\past}^y, K_{\future}^u, K_{\future}^y)$ as kernel matrices of the ``offline'' training data generated by \eqref{eq.NL.state-space.phi}. By splitting the kernel vector of the ``online'' testing sequence $\Phi_{0:2L-1}$ into two parts (past and future), we test whether one can interpolate the kernel matrices of past and future using the same coefficients, where the initial length-$L$ segment of the testing sequence specifies the initial condition for the subsequent length-$L$ segment. A similar observation for LTI systems on the specification of initial condition has been given in \cite[Proposition 1]{markovsky2008data}. In the proof of Theorem~\ref{thm.NL.FL}, we further show that the trajectory passes through a state $\hat{x}_L$ at time $L$ that is a linear combination of states from the training data, i.e., $\hat{x}_L = X_\future \xi$ for some $\xi$.

In light of Theorem \ref{thm.NL.FL}, we can compute the predictor $\hat{y}_{L:2L-1}$ via 
\begin{align*}
\hat{y}_{L:2L-1} 
= (Y_\future^{\trn})^\dagger k_{\future}^y
= (Y_\future^{\trn})^\dagger {K_{\future}^y} \xi,
\end{align*}
where $\xi$ is the solution of the following equation,
\begin{align*}
\begin{bmatrix} k_{\past}^u \\ k_{\past}^y \\ k_{\future}^u \end{bmatrix} 
= \begin{bmatrix} {K_{\past}^u} \\ {K_{\past}^y} \\ {K_{\future}^u} \\ \end{bmatrix} \xi. 
\end{align*}

The interpolation-based method in Section \ref{sec.interpolate} characterizes the future output ${Y_{\future}}$ using kernelized offline data to make predictions regarding the online data. On the other hand, the realization-based method aims to capture the output $y_{t+L-1}$ as linear functions of states $X_{\future}$ and kernelized input vectors $\phi(u_t),\dots,\phi(u_{t+L-1})$. The preference between the two approaches depends on the memory length $L$ and the state dimension $n$. E.g., in the regime $L \gg n$, the states serve as an efficient representation of the system's history.

\section{Conclusion}
\label{sec.conclusion}

In this paper,  we have put forward a behavioral framework for modeling a class of nonlinear systems in a vector-valued RKHS. This formulation is rich enough to cover LTI systems as well as nonlinear systems modeled by Volterra series, autoregressive models based on Volterra series, and Hammerstein-type state-space models. Using this framework, we have analyzed two methods for data-driven modeling of such systems,  minimum-norm interpolation and subspace identification. We have clarified the role of various structural assumptions on the system, the data (both offline and online), and the vector-valued RKHS that represents the nonlinear aspects of the system. More broadly, this work expands the scope of behavioral systems theory to nonlinear systems. In doing so, it reinforces the conceptual shift underlying data-driven control: rather than aiming to recover the state evolution, one can directly actualize the desired behaviors using observed trajectories.

\appendices

\section{Proof of Lemma~\ref{lemma.LVW}}
\label{ap.LVW}

Let $\|\cdot\|_\Vcal$ and $\|\cdot\|_\Wcal$ be the Hilbert-space norms on $\Vcal$ and $\Wcal$ induced by their respective inner products. Consider the evaluation functional $\delta_{v}:\Lcal(\Vcal,\Wcal) \to \Wcal$ given by $\delta_{v}(A) = Av$. It is bounded since
\begin{align*}
\begin{aligned}
\vnorm{\delta_{v}(A)}_{\Wcal}
=& \vnorm{A v}_{\Wcal}\\
\leq& \vnorm{A}_{\Lcal(\Vcal,\Wcal)} \vnorm{v}_{\Vcal} \\
\leq& \vnorm{A}_{\Hcal} \vnorm{v}_{\Vcal},
\end{aligned}
\end{align*}
where the last step follows from the relation
\begin{align*}
\vnorm{A}_{\Lcal(\Vcal,\Wcal)} &= \sup_{v \in \Vcal,\, \vnorm{v}_\Vcal = 1} \sqrt{\langle Av, Av \rangle_\Wcal} \\
&= \sup_{v \in \Vcal,\, \vnorm{v}_\Vcal = 1} \sqrt{\langle v, A^*A v \rangle_\Vcal} \\
&\leq \| A \|_\Hcal.
\end{align*}
Hence, $\Hcal$ is an RKHS. We next show that $\kappa\left(v, v' \right)
= \langle v, v' \rangle_{\Vcal} I_{\Wcal}$ is the reproducing kernel. It is obvious that $\kappa$ is of the positive type. To show it satisfies the reproducing property, notice that for any $v \in \Vcal$ and $w \in \Wcal$ we have
\begin{align*}
\begin{aligned}
\left \langle  A v, w\right \rangle_{\Wcal} 
= \Tr\left( A^* (w \otimes v) \right) \\
= \left \langle A, w \otimes v \right \rangle_{\Hcal},
\end{aligned}
\end{align*}
where $w \otimes v \in \Hcal$ is the rank-one linear operator $(w \otimes v)v' = \langle v,v' \rangle_\Vcal w$.

On the other hand, for $w \in \Wcal$, we have
\begin{align*}
\begin{aligned}
\kappa\left(v', v \right)  w
= \langle v',v \rangle_\Vcal w
= (w \otimes v') v.
\end{aligned}
\end{align*}
Hence, we have
\begin{align*}
\left \langle  A, \kappa(\cdot,v) w \right \rangle_{\Hcal} 
=\left \langle  Av, w\right \rangle_{\Wcal}  .   
\end{align*}
That is, $\kappa$ satisfies the reproducing property. 

\section{Proofs For Results in Section \ref{sec.interpolate}}

\subsection{Proof of Theorem \ref{theorem.NL.<=}}
Let $c_0,\cdots, c_{T-1} \in \Rset$ be given. For each $j=0,\cdots, {T-1}$, the relation
\begin{align*}
\langle y_{j^+}^\data, v \rangle_{\Ycal} 
=  \langle f, \kappa_\Zcal(\cdot,z_j^\data) v  \rangle_{\Hcal_\Zcal}
\end{align*}
holds for all $v \in \Ycal$. Multiplying both sides by $c_j$ and summing over $j = 0,\cdots,{T-1}$, we have
\begin{align*}
\begin{aligned}
\left \langle \sum_{j=0}^{T-1}c_j y_{j^+}^\data, v \right \rangle_{\Ycal} 
=&\left \langle f, \sum_{j=0}^{T-1}c_j \kappa_\Zcal(\cdot,z_j^\data) v \right \rangle_{\Hcal_\Zcal}\\
=&\left \langle \sum_{j=0}^{T-1}c_j f(z_j^\data), v \right \rangle_{\Hcal_\Zcal},
\end{aligned}
\end{align*}
for all $v \in \Ycal$, where the last line follows from the reproducing property.
Hence, the pair $\left(\sum_{j=0}^{T-1}c_j \kappa_Z(\cdot,z_j^\data),\sum_{j=0}^{T-1}c_j y_{j^+}^\data \right)$ satisfies Eq.~\eqref{eq.NL.AR.f} as claimed.

\subsection{Proof of Lemma \ref{lemma.Sigma.t}}
Define the following subspace $\Hcal_N$ of $\Hcal_\Zcal$:
\begin{align}
\begin{aligned}
\Hcal_N &:= {\rm span}\left\{\kappa_\Zcal(\cdot,z_i)v, \ i = 0,\cdots,N-1, \ v \in \Ycal \right\}\\
&= {{\rm range}(\Phi_N)}.
\end{aligned}
\label{eq.H.t-1}
\end{align}
Let $\Pi_N$ denote the orthogonal projection onto $\Hcal_N$ and notice that $\Pi_N =  \Phi_N \left(\Phi_N^*\Phi_N\right)^\dagger \Phi_N^*$. Using this in the definition of $\Sigma_N(z)$ in \eqref{def.Sigma}, we get
\begin{align*}
\begin{aligned}
\Sigma_N(z) 
&= \kappa_\Zcal(z,z) - k_N(z)K_N^\dagger {k}_N(z)^*\\
&= \kappa_\Zcal(\cdot,z)^*\kappa_\Zcal(\cdot,z) \\
&\quad -\kappa_\Zcal(\cdot,z)^* \Phi_N \left(\Phi_N^*\Phi_N\right)^\dagger \left(\kappa_\Zcal(\cdot,z)^* \Phi_N \right)^* \\
&=\kappa_\Zcal(\cdot,z)^* \big( I -  \Phi_N (\Phi_N^*\Phi_N)^\dagger \Phi_N^*  \big) \kappa_\Zcal(\cdot,z)\\
&=\kappa_\Zcal(\cdot,z)^* (I-\Pi_N) \kappa_\Zcal(\cdot,z).
\end{aligned}
\end{align*}
Hence, for any $v \in \Rset^p$, 
\begin{align*}
\begin{aligned}
v^\trn \Sigma_N(z) v
=& v^\trn \kappa_\Zcal(\cdot,z)^* (I-\Pi_N) \kappa_\Zcal(\cdot,z) v\\
=& \vnorm{(I-\Pi_N) \kappa_\Zcal(\cdot,z) v}_{\Hcal_\Zcal}.
\end{aligned}
\end{align*}
Hence, $\Sigma_N(z)$ is  positive semi-definite. In particular, $\Sigma_N(z) = 0$ iff $(I-\Pi_N )\kappa_\Zcal(\cdot,z) v = 0$ for all $v \in \Rset^p$, that is, if $\kappa_\Zcal(\cdot,z)v \in \Hcal_N = {\rm range}(\Phi_N)$.
\subsection{Proof of Lemma \ref{lemma.liang.vRKHS}}

Since $f_{N+1}$ interpolates $\{(z_i, y_i)\}^{N-1}_{i=0}$, we have for all $v \in \Ycal$ and $i = 0,\cdots, N-1$,
\begin{align*}
\begin{aligned}
\left \langle f_{N+1}, \kappa_\Zcal(\cdot,z_i)v \right \rangle_{\Hcal_\Zcal} 
=\left \langle f_{N+1}(z_i),v \right \rangle_{\Ycal} 
= \left \langle y_i , v \right \rangle_{\Ycal}.
\end{aligned}
\end{align*}
On the other hand,  for $i = 0,\cdots, N-1$ and all $v \in \Ycal$, 
\begin{align*}
\begin{aligned}
&\left \langle f_{N+1}, \kappa_\Zcal(\cdot,z_i)v \right \rangle_{\Hcal_\Zcal} \\
&= \left \langle \Pi_N[f_{N+1}]+ \Pi_N^\perp[f_{N+1}], \kappa_\Zcal(\cdot,z_i)v \right \rangle_{\Hcal_\Zcal} \\
&= \left \langle \Pi_N[f_{N+1}],\kappa_\Zcal(\cdot,z_i)v \right \rangle_{\Hcal_\Zcal},
\end{aligned}
\end{align*}
where $\Pi^\perp_{N} = I - \Pi_N$ and $\Pi_N$ is the orthogonal projection onto $\Hcal_{N}$.  Since $v$ is arbitrary,  $\Pi_N[f_{N+1}](z_i) = y_i$ for $i = 0,\cdots, N-1$, i.e., both $f_{N}$ and $\Pi_N[f_{N+1}]$  interpolate $\{(z_i,y_i)\}_{i=0}^{N-1}$. Since $f_{N}$ is the unique minimum-norm solution in $\Hcal_{N}$, it must be the case that $f_{N} = \Pi_N[f_{N+1}]$. Hence,  
\begin{align*}
f_{N+1} - f_{N}
=f_{N+1} - \Pi_N[f_{N+1}] \in \Hcal_{N}^\perp.
\end{align*}
That is, there exists some $\xi_N \in \Ycal$ such that
\begin{align}
\begin{aligned}
f_{N+1} - f_{N}
= \Pi_N^\perp[\kappa_\Zcal(\cdot,z_N) \xi_N ].
\end{aligned}
\label{eq.liang.orthogonal}
\end{align}
We now determine $\xi_N$. By the reproducing property, we have
\begin{align}
\begin{aligned}
&\left \langle f_{N+1}(z_N), v \right \rangle_{\Ycal} \\
&=\left \langle f_{N+1}, \kappa_\Zcal(\cdot,z_N)v \right \rangle_{\Hcal_\Zcal} \\
&= \left \langle f_{N} + \Pi_N^\perp[\kappa_\Zcal(\cdot,z_N) \xi_N ], \kappa_\Zcal(\cdot,z_N)v \right \rangle_{\Hcal_Z} \\
&= \left \langle f_{N}, \kappa_\Zcal(\cdot,z_N)v \right \rangle_{\Hcal_\Zcal} \\
&\quad + \left \langle\Pi_N^\perp[\kappa_\Zcal(\cdot,z_N) \xi_N ], \kappa_\Zcal(\cdot,z_N)v \right \rangle_{\Hcal_\Zcal}.
\end{aligned}
\label{eq.liang.1}
\end{align}
For the last line, applying the reproducing property again, we can write the first term as $ \langle f_{N}, \kappa_\Zcal(\cdot,z_N)v  \rangle_{\Hcal_\Zcal}= \langle f_{N}(z_N), v  \rangle_{\Ycal}$. For the second term, by orthogonality we have
\begin{align*}
\begin{aligned}
&\left \langle\Pi_N^\perp[\kappa_\Zcal(\cdot,z_N) \xi_N ], \kappa_\Zcal(\cdot,z_N)v \right \rangle_{\Hcal_\Zcal} \\
&= \left \langle \Pi_N^\perp[\kappa_\Zcal(\cdot,z_N) \xi_t ], 
 \Pi_N^\perp[\kappa_\Zcal(\cdot,z_N) v ] \right \rangle_{\Hcal_\Zcal} \\
&= \langle \xi_N, \kappa_\Zcal(z_N,z_N) v \rangle_{\Hcal_\Zcal} \\
&\quad- \left \langle \Pi_N[\kappa_\Zcal(\cdot,z_N) \xi_N], \Pi_N[\kappa_\Zcal(\cdot,z_N) v ] \right \rangle_{\Hcal_\Zcal},
\end{aligned}
\end{align*}
where we have used the fact that, since $\Pi_N$ is an orthogonal projection, 
\begin{align*}
\langle \Pi_N h, \Pi_N h' \rangle_{\Hcal_\Zcal} = \langle h, \Pi_N h' \rangle_{\Hcal_\Zcal} = \langle \Pi_N h, h' \rangle_{\Hcal_\Zcal}
\end{align*}
for all $h,h' \in \Hcal_\Zcal$. Since $\Pi_N$ is an orthogonal projection onto $\Hcal_{N}$,  we have
\begin{align*}
\Pi_N[\kappa_\Zcal(\cdot,z_N) \xi_N]
= \sum_{j=0}^{N-1} \kappa_\Zcal(\cdot,z_j) \alpha_j,
\end{align*}
and
\begin{align*}
\Pi_N[\kappa_\Zcal(\cdot,z_N) v]
= \sum_{j=0}^{N-1} \kappa_\Zcal(\cdot,z_j) \beta_j,
\end{align*}
where $\bar{\alpha} = (\alpha_0, \cdots \alpha_{N-1}) \in \Ycal^{N-1}$ and $\bar{\beta} = (\beta_0,\dots,\beta_{N-1}) \in \Ycal^{N-1}$ are  the solutions of $K_{N} \bar{\alpha} =  k_{N}(z_N) \xi_N$ and $K_{N}\bar{\beta} = k_{N}(z_N)v$. Therefore,
\begin{align*}
\begin{aligned}
&\left \langle \Pi_N[\kappa_\Zcal(\cdot,z_N) \xi_N ], \Pi_N[\kappa_\Zcal(\cdot,z_N) v ] \right \rangle_{\Hcal_\Zcal} \\
&= \left \langle \sum_{j=0}^{N-1} \kappa_\Zcal(\cdot,z_j) \alpha_j, \sum_{j=0}^{N-1} \kappa_\Zcal(\cdot,z_j) \beta_j \right \rangle_{\Hcal_\Zcal}\\
&= \left \langle \bar{\alpha}, K_{N} \bar{\beta}\right \rangle_{\Ycal^{\oplus (N-1)}} \\
&= \left \langle  K_{N}^\dagger k_{N}(z_N)\xi_N, K_{N}K_{N}^\dagger  k_{N}(z_N) v \right \rangle_{\Ycal^{\oplus (N-1)}}\\
&= \left \langle \xi_N, k^*_{N}(z_N) K_{N}^\dagger  k_{N}(z_N) v \right \rangle_{\Ycal},
\end{aligned}
\end{align*}
where the last line holds since $K_{N}^\dagger K_{N} K_{N}^\dagger = K_{N}^\dagger$.

Putting everything together and using the definition of $\Sigma_N(z_N)$, we arrive at
\begin{align*}
\begin{aligned}
&\left \langle \Pi_N^\perp[\kappa_\Zcal(\cdot,z_N) \xi_N ], 
\Pi_N^\perp[\kappa_\Zcal(\cdot,z_N) v ] \right \rangle_{\Hcal_\Zcal}  \\
&= \left \langle \xi_N, \kappa_\Zcal(z_N,z_N) v \right \rangle_{\Ycal} 
- \left \langle \xi_N, k^*_{N}(z_N) K_{N}^\dagger  k_{N}(z_N) v\right \rangle_{\Ycal}\\
&= \left \langle \xi_N, \Sigma_{N}(z_N) v \right \rangle_{\Ycal}.
\end{aligned}
\end{align*}
Plugging the above relation back into \eqref{eq.liang.1} and using the fact that $v \in \Ycal$ is arbitrary and that $\Sigma_{N}(z_N)$ is self-adjoint,  we see that $\xi_N$ is determined by the relation
\begin{align*}
f_{N+1}(z_N) = f_{N}(z_N)+ \Sigma_{N}(z_N) \xi_N.
\end{align*}
As $f_{N+1}$ interpolates $\left(z_N,y_{N^+}\right)$, i.e., $y_{N^+} = f_{N+1}(z_N)$, we can further write
\begin{align}
\Sigma_{N}(z_N) \xi_N =  y_{N^+} -  f_{N}(z_N).
\label{eq.liang.ct}
\end{align}
Hence, when $\Sigma_{N}(z_N)=0$, we will have $f_{N}(z_N) = f_{N+1}(z_N) = y_{N^+}$.

Next,  applying the Pythagorean theorem in  \eqref{eq.liang.orthogonal}, we have
\begin{align*}
\begin{aligned}
\vnorm{f_{N+1}}_{\Hcal_\Zcal}^2
=& \vnorm{f_{N}}_{\Hcal_\Zcal}^2
+ \vnorm{\Pi_N^\perp[\kappa_\Zcal(\cdot,z_N) \xi_t ]}_{\Hcal_\Zcal}^2 \\
=& \vnorm{f_{N}}_{\Hcal_\Zcal}^2
+  \left \langle \xi_N, \Sigma_{N}(z_N) \xi_N\right \rangle_{\Ycal},
\end{aligned}
\end{align*}
where the last line follows from the definition of $\Sigma_{N}$. 
When $\Sigma_{N}(z_N) \succ 0$, from \eqref{eq.liang.ct} we have
\begin{align*}
\begin{aligned}
&\vnorm{f_{N+1}}_{\Hcal_\Zcal}^2 - \vnorm{f_{N}}_{\Hcal_\Zcal}^2\\
&=\left \langle \Sigma_{N}^{-1}(z_N) \left( y_{N^+} -  f_{N}(z_N)\right), y_{N^+} -  f_{N}(z_N)\right \rangle_{\Ycal}\\
&= \vnorm{\Sigma_{N}^{-1/2}(z_N) \left(y_{N^+} -  f_{N}(z_N)\right)}^2_{\Ycal}.
\end{aligned}
\end{align*}
This concludes the proof.
\section{Proofs For Results in Section \ref{sec.subspace}}

\subsection{Proof of Theorem \ref{thm.NL.FL}}

In our model \eqref{eq.NL.state-space.phi}, we can view $v = \phi(u)$ as the $q$-dimensional input to the LTI system parametrized by $(A,B,C,D)$. Consequently, the argument in the proof in \cite[Theorem 1]{van2020willems} extends directly to our setting and guarantees that the matrix
\begin{align*}
\Mcal := \begin{bmatrix}  {U_\past^\phi} \\ {U_\future^\phi}  \\X_0 \end{bmatrix} \in \Rset^{(2Lq+n) \times (T-2L+1)}
\end{align*} 
has full row rank.

$(\Rightarrow)$: Suppose that $\begin{bmatrix} {\Phi}_{0:2L-1} \\ {y}_{0:2L-1} \end{bmatrix}$ is a valid input-output trajectory of \eqref{eq.NL.state-space.phi}. Then there exists an initial state $\hat{x}_0$ such that
\begin{align*}
{y}_{0:L-1} =   \Ocal_L \hat{x}_0+ \Tcal_L {\Phi}_{0:L-1}.
\end{align*}
Since $\Mcal$ has full row rank, for any $\begin{bmatrix} {\Phi}_{0:L-1} \\ {\Phi}_{L:2L-1} \\ \hat{x}_0
\end{bmatrix} \in \Rset^{2Lq} \oplus \Rset^n$, there exists some $\xi \in \Rset^{T-2L+1}$ such that 
\begin{align*}
\begin{bmatrix} {\Phi}_{0:L-1} \\ {\Phi}_{L:2L-1} \\ \hat{x}_0
\end{bmatrix}  = \Mcal \xi .
\end{align*}
Thus, we have
\begin{align*}
\begin{aligned}
\begin{bmatrix}
{\Phi}_{0:L-1} \\  {y}_{0:L-1}   
\end{bmatrix}
=&\begin{bmatrix} I & {0} \\ \Tcal_L & \Ocal_L \end{bmatrix}
\begin{bmatrix}
{\Phi}_{0:L-1} \\ \hat{x}_0
\end{bmatrix} \\
=& \begin{bmatrix} I & {0} \\ \Tcal_L & \Ocal_L \end{bmatrix}
\begin{bmatrix} {U_\past^\phi} \\ X_0 \end{bmatrix} \xi \\
=& \begin{bmatrix} {U_\past^\phi} \\ {Y_\past} \end{bmatrix} \xi.
\end{aligned}
\end{align*}
Moreover, from the dynamics \eqref{eq.NL.state-space.phi}, we have
\begin{align}
X_\future = A^L X_0 + \Delta_L {U_\past^\phi}.
\label{eq.Xf}
\end{align}
Hence, at time $t=L$, 
\begin{align*}
\begin{aligned}
\hat{x}_L =& A^L \hat{x}_0+ \Delta_L {\Phi}_{0:L-1}\\
=& \left( A^L X_0 + \Delta_L {U_\past^\phi}\right)\xi \\ 
=& X_\future \xi.
\end{aligned}
\end{align*}
Thus, starting from $\hat{x}_L= X_\future \xi$ with input sequence ${\Phi}_{L:2L-1} = {U_\future^\phi} \xi$, we have
\begin{align*}
\begin{aligned}
\begin{bmatrix}
{\Phi}_{L:2L-1} \\ {y}_{L:2L-1}   
\end{bmatrix}
=& \begin{bmatrix} I & {0} \\ \Tcal_L & \Ocal_L \end{bmatrix}
\begin{bmatrix}
{\Phi}_{L:2L-1} \\ \hat{x}_L
\end{bmatrix}\\
=& \begin{bmatrix} I & {0} \\ \Tcal_L & \Ocal_L \end{bmatrix}
\begin{bmatrix} {U_\future^\phi} \\ X_\future \end{bmatrix} \xi \\
=& \begin{bmatrix} {U_\future^\phi} \\ {Y_\future} \end{bmatrix} \xi.
\end{aligned}
\end{align*}

$(\Leftarrow)$ Since $\begin{bmatrix} {u}_{0:L-1} \\ {y}_{0:L-1} \end{bmatrix} 
= \begin{bmatrix} {U_\past^\phi} \\ {Y_\past} \end{bmatrix} \xi$,
and $\begin{bmatrix} {U_\past^\phi} \\ {Y_\past} \end{bmatrix}$ is constrained by \eqref{eq.NL.state-space.phi}, we have
\begin{align*}
\begin{bmatrix} {\Phi}_{0:L-1} \\ {y}_{0:L-1} \end{bmatrix} 
= \begin{bmatrix} {U_\past^\phi} \\ {Y_\past} \end{bmatrix} \xi
= \begin{bmatrix} I & {0} \\ \Tcal_L & \Ocal_L \end{bmatrix}
\begin{bmatrix} {U_\past^\phi} \\ X_0 \end{bmatrix} \xi.
\end{align*}
Hence,
\begin{align*}
{y}_{0:L-1} = {Y_\past} \xi 
= \Ocal_L X_0 \xi + \Tcal_L {U_\past^\phi} \xi
\end{align*}
Since ${\Phi}_{0:L-1}={U_\past^\phi} \xi$, we conclude that ${y}_{0:L-1}$ is the output from $\hat{x}_0:= X_0 \xi $ with input ${u}_{0:L-1}$. As $\left({\Phi}_{0:L-1},{y}_{0:L-1}\right)$ is a valid input-output trajectory, we can use \eqref{eq.NL.state-space.phi} and $X_\future$ defined in \eqref{eq.Xf} to obtain the state vector at time $L$ as
\begin{align}
\hat{x}_L = A^L \hat{x}_0+ \Delta_L {\Phi}_{0:L-1}
= A^L X_0 \xi + \Delta_L {U_\past^\phi} \xi
= X_\future \xi.
\label{eq.xhat_L}
\end{align}

For the second segment, using the assumptions and the matrix input-output relation for ${U_\future^\phi}$, ${Y_\future}$, we have
\begin{align*}
\begin{bmatrix} {\Phi}_{L:2L-1} \\ {y}_{L:2L-1} \end{bmatrix}
= \begin{bmatrix} {U_\future^\phi} \\ {Y_\future} \end{bmatrix} \xi
= \begin{bmatrix} I & {0} \\ \Tcal_L & \Ocal_L \end{bmatrix}
\begin{bmatrix} {U_\future^\phi} \\ X_\future \end{bmatrix} \xi.
\end{align*}
Thus, we have  
\begin{align*}
\begin{aligned}
{y}_{L:2L-1} 
&= {Y_\future} \xi \\
&= \Ocal_L X_\future \xi + \Tcal_L {U_\future^\phi} \xi \\
&= \Ocal_L \hat{x}_L + \Tcal_L {\Phi}_{L:2L-1},
\end{aligned}
\end{align*}
where the last equality follows from \eqref{eq.xhat_L}. 
Therefore, we conclude that ${y}_{L:2L-1}$ is the output from $\hat{x}_L= X_\future \xi $ with input ${\Phi}_{L:2L-1}$.

Putting everything together, we have
\begin{align*}
\begin{aligned}
\begin{bmatrix} {\Phi}_{0:L-1} \\ {y}_{0:L-1} \end{bmatrix} 
= \begin{bmatrix} {U_{\past}^\phi} \\ {Y_{\past}} \end{bmatrix} \xi , \quad
\begin{bmatrix}
{\Phi}_{L:2L-1} \\ {y} _{L:2L-1} \end{bmatrix} =\begin{bmatrix} {U_{\future}^\phi}\\ {Y_{\future}} \end{bmatrix} \xi.
\end{aligned}
\end{align*}
Multiplying both sides of the first equation by $\begin{bmatrix} {U_{\past}^\phi}^* & 0 \\ 0 & Y_{\past}^*  \end{bmatrix}$ and the second equation by $\begin{bmatrix} {U_{\future}^\phi}^* & 0 \\ 0 & Y_{\future}^* \end{bmatrix}$ proves the claim.

\bibliographystyle{IEEEtran}
\small
\bibliography{ref_behavior,ref_RKHS}

\end{document}